\documentclass[aps,prl,showpacs,onecolumn,floatfix,amsmath]{revtex4}
\usepackage{graphicx}
\usepackage{epstopdf}
\usepackage{bm}

\begin{document}
\title[Optimal control of population and coherence in three-level $\Lambda$~systems]{Optimal control of population and coherence in three-level $\Lambda$~systems}

\author{Praveen~Kumar, Svetlana~A.~Malinovskaya, Vladimir~S.~Malinovsky\footnote{Present address:
ARL, 2800 Powder Mill Road, Adelphi, MD 20783}
}

\address{Department of Physics and Engineering Physics, Stevens Institute of Technology, Hoboken, New Jersey 07030}

%\ead{Vladimir.Malinovsky@stevens.edu}

\begin{abstract}
Optimal control theory implementations for an efficient population transfer and creation of a maximum coherence in three-level system
are considered. We demonstrate that the half-STIRAP (stimulated Raman adiabatic passage) scheme for creation of the maximum Raman coherence is the optimal solution according to the optimal control theory.
We also present a comparative study of several implementations of optimal control theory applied to the complete population transfer and creation of the maximum coherence. Performance of the conjugate gradient method, the Zhu-Rabitz method and the Krotov method has been analyzed.
\end{abstract}
%\submitto{\JPB}
\maketitle

\section{Introduction}
Efficient and selective transfer of population in quantum systems provides an essential tool for a variety of applications in physical chemistry, laser spectroscopy, quantum optics, and quantum information processing~\cite{rice,shapiro,eberly,berman,nielsen}. Some of the population transfer techniques make use of the Rabi oscillations when the transfer efficiency is controlled by pulse area of the external fields~\cite{shore}.
Sensitivity of the final population distribution in the system to the field parameters is sometimes considered as a drawback of these methods.
Techniques utilizing adiabatic passage solution in the system dynamics are substantially more robust against moderate variations in the interaction parameters. Stimulated Raman adiabatic
passage (STIRAP) is one example of such methods~\cite{bergmann,kral}.

In a three-level $\Lambda$ system, STIRAP provides a robust scheme for transferring population by using two strong pulses, called the pump and the Stokes~\cite{bergmann,kral}. When the pump-Stokes pulse sequence is arranged in a counterintuitive order (the Stokes pulse precedes the pump pulse) the system dynamics takes place within a so-called dark state (a coherent superposition of the initial and the target states). The adiabatic change of the field amplitudes guaranties nearly 100\% efficiency of population transfer to the target state while the intermediate state population is negligibly small during whole time evolution. There are a few modifications of the STIRAP scheme for more complex configurations of quantum systems~\cite{bergmann,kral,shapiro1,vardi,nakajima,yatsenko,vitanov1,malinovsky1,Kis-1}.

Search for an efficient scheme of population transfer is also one of the major goals of optimal control theory (OCT), a powerful and sometimes mathematically very sophisticated technique~\cite{zhu,somloi,tannor,palao,palao1}.
The OCT algorithm has been applied successfully to a broad variety of physical and chemical systems \cite{vivie-riedle,kumar,branderhorst,ren,bartana,tosner,zare,levis,Kis-2,kumarS}. It intrinsically accounts for the quantum-mechanical interference between many pathways connecting initial and target state in a quantum system.
There were some interesting arguments in trying to find a link between OCT and STIRAP. Despite initial negative prognosis~\cite{band}, it was demonstrated recently that STIRAP type solution can emerge automatically from global OCT method~\cite{malinovsky}. Besides, STIRAP scheme has been obtained successfully using a local optimal control theory~\cite{malinovsky1}.

In this work we review a performance of the optimal control theory algorithms~\cite{rice,shapiro,Kis-2,balint-kurti,kosloff,peirce,shi} applied  for the population transfer and explore their potential to maximize the Raman coherence in the three-level $\Lambda$ system. Our motivation for
maximizing coherence is related to the recent developments in coherent anti-Stokes Raman (CARS) microscopy and remote detection using intense femto-second laser setups~\cite{xie,scully,silberberg,dantus,malinovskaya,malinovskaya1}.

The paper is organized as follows. In section II, basic equations of the optimal control theory are developed using the method of variational
calculus. Field equations are derived using the penalty on the energy of the control field. A time dependent penalty function is used which
ensures experimentally feasible profile of the laser pulses. A second penalty function is introduced to minimize the
population of the excited intermediate state throughout time evolution of the system. In section III, we apply the OCT formalism to a three-level $\Lambda$ system and analyze solutions of the OCT equations for population transfer and a maximum Raman coherence applying
different optimization strategies: the conjugate gradient method~\cite{Kis-2,balint-kurti}, the Zhu-Rabitz method~\cite{zhu} and the Krotov method~\cite{somloi,tannor,palao,palao1}. Section IV is the conclusion.

\section{General equations of Optimal Control Theory}
The OCT is based on the definition of a cost functional $K$ which must have an optimal value when the desired transformation of the wave function is successfully achieved by the control laser field $\epsilon(t)$. An optimal solution requires that the system wave function at a final time, $| \psi (T)\rangle $, should be as close as possible to the target wave function, $ |\phi (T)\rangle $. That is the overlap $ \vert \langle \psi (T) \vert \phi (T) \rangle \vert^2$ is maximal at final time $T$.

In order to derive control equation for the field and obtain realistic field amplitude we minimize the energy fluency of the field. Another requirement is that the population of the excited intermediate states has to be minimal throughout the transfer process \cite{somloi,palao,malinovsky}, that can be done by defining a projection operator  $\hat{P}=|\psi_{int}(t)\rangle\langle \psi_{int}(t)|=\sum_k |k\rangle\langle k|$, where $|k\rangle$ is the eigenket of the unwanted intermediate states. There is an additional constraint that should be taken into account when we maximize or minimize the cost functional $K$: the wave function $|\psi_{}(t)\rangle $ must satisfy the Schr\"odinger equation.

These requirements lead to a complete cost functional of the form
\begin{widetext}
%\begin{eqnarray}\label{eq-K}
\begin{equation}\label{eq-K}
K = \big\vert \langle \psi(T) \vert \phi(T) \rangle \big\vert^2
- \alpha(t) \int_0^T dt \big[ \epsilon(t) - \epsilon_r(t) \big]^2
- \beta \int_0^T dt \langle \psi(t)|\hat{P} | \psi (t) \rangle
% \nonumber \\
-  2 {\rm Re} \Big[
\int_0^T dt \Big\langle \chi (t) \Big\vert \frac{\partial}{\partial t} + \frac{i}{\hbar} \hat{H} \Big\vert \psi (t) \Big\rangle
 \Big].
%\end{eqnarray}
\end{equation}
\end{widetext}
The factor $\alpha (t)=\alpha_0/s(t)$ is a time dependent penalty function that determines the shape function $s(t)$,
$\alpha_0$ is a constant which should be determined due to the significance of the field energy value. The main purpose of $s(t)$ is to turn the field on and off smoothly to ensure feasibility of experimental implementation of the optimal laser pulse. $\epsilon_r(t)$ denotes a
reference field and $\beta$ is a penalty parameter for the total population of the intermediate state (or state manifold). The function $|\chi (t)\rangle$ can be regarded as a Lagrange multiplier introduced to assure satisfaction of the Schr\"odinger equation.

Each of the terms in Eq.~(\ref{eq-K}) depends explicitly or implicitly on the unknown driving field. The goal is to determine an optimal field  $\epsilon(t)$ for which the cost functional has an extremum. Taking variations  of the cost functional with respect to $|\chi (t)\rangle$, $|\psi (t)\rangle$ and $\epsilon(t)$ we find the following set of equations

\begin{equation}\label{initial}
 i \hbar \frac{\partial}{\partial t} |\psi (t)\rangle = \hat{H} | \psi (t) \rangle \, ,
\end{equation}
\begin{equation}\label{lagrange}
\frac{\partial}{\partial t} |\chi(t)\rangle = - \frac{i}{\hbar} \hat{H} |\chi(t)\rangle + \beta \hat{P} |\psi(t)\rangle \, ,
\end{equation}
\begin{eqnarray}\label{field}
\epsilon(t) = \epsilon_r (t) + \frac{1}{ \alpha(t) \hbar} {\rm Im}
\Big\langle \chi(t) \Big\vert \frac{\partial\hat{H}}{\partial \epsilon(t)} \Big\vert \psi(t) \Big\rangle.
\end{eqnarray}

Variation of the cost functional with respect to $|\psi(T)\rangle$ gives the initial condition for the Lagrange multiplier
\begin{equation}
|\chi(T)\rangle = \langle \psi(T) \vert \phi (T) \rangle |\phi(T)\rangle \, .
\end{equation}

Equations (\ref{initial}) and (\ref{lagrange}) determine the time evolution of the wave function and Lagrange multiplier which are used in Eq.~(\ref{field}) to find the optimal field, $\epsilon(t)$, maximizing  the overlap $ \vert \langle \psi(T) \vert \phi(T) \rangle \vert^2$ in Eq.~(\ref{eq-K}).

\section{Optimal control of three-level $\Lambda$ system dynamics}
\subsection{General equations}

To demonstrate an application of the general optimization formalism outlined in the previous section and to test the performance of various implementations of the OCT we consider two optimal control problems. First, we reexamine the population transfer from the initially occupied level $|1\rangle$ to the level $|3\rangle$. In the second problem, we utilize the OCT to create a maximum Raman coherence, in other words, the 50/50 coherent superposition of states $|1\rangle$ and $|3\rangle$.

\begin{figure}
\centerline{\includegraphics[angle=-90,width=6cm]{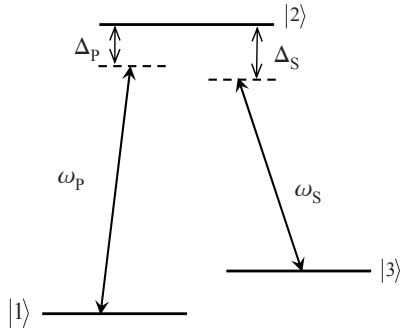}}
\caption{Schematic of a three-level system.}\label{scheme1}
\end{figure}

First, we consider the population transfer in a generic three-level $\Lambda$ system exited by a pump-Stokes pulse sequence, see Fig.~\ref{scheme1}. We address the so-called nonimpulsive excitation when the pump pulse interacts with $\vert 1 \rangle - \vert 2 \rangle$ transition while the Stokes pulse controls the transition $\vert 2 \rangle - \vert 3 \rangle$. We assume that all relaxation times in the system are much longer than the pulse duration so that the dynamics of an  arbitrary wave function
\begin{equation}
|\psi(t)\rangle = a_1(t) \vert 1 \rangle + a_2(t) \vert 2 \rangle + a_3(t)
\vert 3 \rangle \, ,
\end{equation}
where $a_{i}(t)$ is the probability amplitude to be in state $|i\rangle$, is governed by the Schr\"odinger equation with the Hamiltonian of the form
\begin{eqnarray}\label{eq2}
{\hat{H}} =
\left(\begin{array}{ccc}
E_{1} & -\mu_{12} \epsilon_{P}^{}(t) & 0 \\
-\mu_{21} \epsilon_{P}^{}(t)  & E_{2} &  -\mu_{23} \epsilon_{S}^{}(t) \\
0 & -\mu_{32} \epsilon_{S}^{}(t) & E_{3} \\
\end{array} \right) \, ,
\end{eqnarray}
here $E_{i}$ is the energy of the $i=1,2,3$ state, $\mu_{12,23}$ are the dipole moments, $\epsilon_{P,S}(t)$ are the pump and Stokes fields.

Now we are ready to solve the optimization problem by applying a general algorithm outlined in the previous section. Varying the cost functional with respect to the Lagrange multiplier vector $b_i(t)$ , the probability
amplitudes vector $a_i(t)$, and the external field $\epsilon_{P,S}(t)$, Eqs. (\ref{initial}), (\ref{lagrange}) and (\ref{field}) takes the following
form:
\begin{equation}\label{eq-a}
i \hbar \frac{\partial a_i(t)}{\partial t} = \hat{H}_{ij} a_j(t),
\end{equation}

\begin{equation}\label{eq-b}
\frac{\partial b_i(t)}{\partial t} = -\frac{i}{\hbar} \hat{H}_{ij} b_j(t) + \beta a_2(t) \delta_{i2},
\end{equation}

\begin{equation}\label{eq-Ep}
\epsilon_{P}(t) = \epsilon_{P}^{r}(t)-\frac{1}{2 \alpha(t)} \cdot {\rm Im} \Big[ b_1^{\star}(t) a_2(t) + b_2^{\star}(t) a_1(t) \Big],
\end{equation}

\begin{equation}\label{eq-Es}
\epsilon_{S}(t) = \epsilon_{S}^{r}(t) -\frac{1}{2 \alpha(t)} \cdot {\rm Im} \Big[ b_2^{\star}(t) a_3(t) + b_3^{\star}(t) a_2(t) \Big] \, ,
\end{equation}
where $\epsilon_{P}^{r}(t)$ and $\epsilon_{S}^{r}(t)$ are the pump and Stokes reference fields.

The set of Eqs.~(\ref{eq-a}) - (\ref{eq-Es}) provides optimal control fields which are based on the solution of the Schr\"odingrer equation with the Hamiltonian in Eq.~(\ref{eq2}). In many cases, we consider the interaction of the three-level system with the external field as a resonant excitation and invoke the Rotating Wave Approximation (RWA) which partially simplifies the solution of the problem and sometimes  facilitates understanding of underlaying physical mechanisms.

In the interaction representation the Hamiltonian can be written in the form
\begin{widetext}
\begin{eqnarray}\label{eq2-interact}
{\hat{H}} = \frac{\hbar}{2}
\left(\begin{array}{ccc}
0 & -W_{P}^{}(t) e^{i\Delta_{P}t}  & 0 \\
-W_{P}^{*}(t) e^{-i\Delta_{P}t}  & 0 &   -W_{S}^{*}(t)e^{-i\Delta_{S}t} \\
0 & -W_{S}^{}(t) e^{i\Delta_{S}t} & 0 \\
\end{array} \right) \, ,
\end{eqnarray}
\end{widetext}
where $W_{P,S}(t)=\Omega_{P,S}(t) (1+e^{-2i \omega_{P,S} t})$, $\Omega_{P,S}(t)=\mu_{12,23} E^0_{P,S}(t)/\hbar$  are the pump and Stokes Rabi frequencies, $E^0_{P,S}(t)$ are the envelope of the pump and Stokes pulses,
$\Delta_{P,S} = \omega_{P,S} - \omega_{21,23}$ are the single photon detunings of the pump and Stokes central frequency $\omega_{P,S}$ from the respective transition frequency $\omega_{21,23}$.

Using the RWA we can replace $W_{P,S}^{}(t)$ in Eq.~(\ref{eq2-interact}) by the corresponding Rabi frequencies, that is we neglect the rapidly oscillating terms. As a result we rewrite the equations for the optimal fields, Eq.~(\ref{eq-Ep}), (\ref{eq-Es}), in terms of the Rabi frequency envelopes of the pump and Stokes fields
\begin{equation}\label{eq-Ep1}
\Omega_P^{ }(t) = \Omega_P^{r}(t) - \frac{1}{2 \alpha(t)} \cdot {\rm Im}
\Big[ b_1^{\star}(t) a_2(t) + b_2^{\star}(t) a_1(t) \Big],
\end{equation}

\begin{equation}\label{eq-Es1}
\Omega_S^{ }(t) = \Omega_S^{r}(t) - \frac{1}{2 \alpha(t)} \cdot {\rm Im}
\Big[ b_2^{\star}(t) a_3(t) + b_3^{\star}(t) a_2(t) \Big],
\end{equation}
where $\Omega_P^{r}(t), \Omega_S^{r}(t)$ are the reference Rabi frequency of the pump and Stokes pulses.

\subsection{Complete population transfer}

To examine OCT implementation we consider three different optimization methods: the conjugate gradient method~\cite{Kis-2,balint-kurti,press}, the Zhu-Rabitz method~\cite{zhu} and the Krotov method~\cite{somloi,tannor}. A detailed description of the numerical  schemes is given in the Appendix. We choose the Gaussian form for the pump and Stokes pulse envelopes as an initial guess
\begin{equation} \label{eq4}
{\Omega}_{P,S}(t) = \Omega_0 \exp \Big( \frac{-(t-t_c)^2}{2\tau_0^2} \Big),
\end{equation}
and our target time is equal to $T=10$ in the units normalized by the time duration $\tau_0$. The reference envelope $\Omega_{P,S}^{r}(t) = 0$ is used unless otherwise stated.

The goal of the first problem at hands is twofold: first, to design the shape and the sequence of the pump and Stokes pulses providing a complete population transfer to state $|3\rangle$; second to suppress population of the intermediate state during excitation process by applying a penalty on the state population. By doing this we also would like to determine the efficiency of the methods mentioned above. For simplicity we restrict our consideration to the exact resonance conditions, $\Delta_{P}=\Delta_{S}=0$.

\begin{widetext}
\begin{figure}[hbt]
\centering
\begin{tabular}{cc}
\includegraphics[width=0.4\textwidth,angle=0]{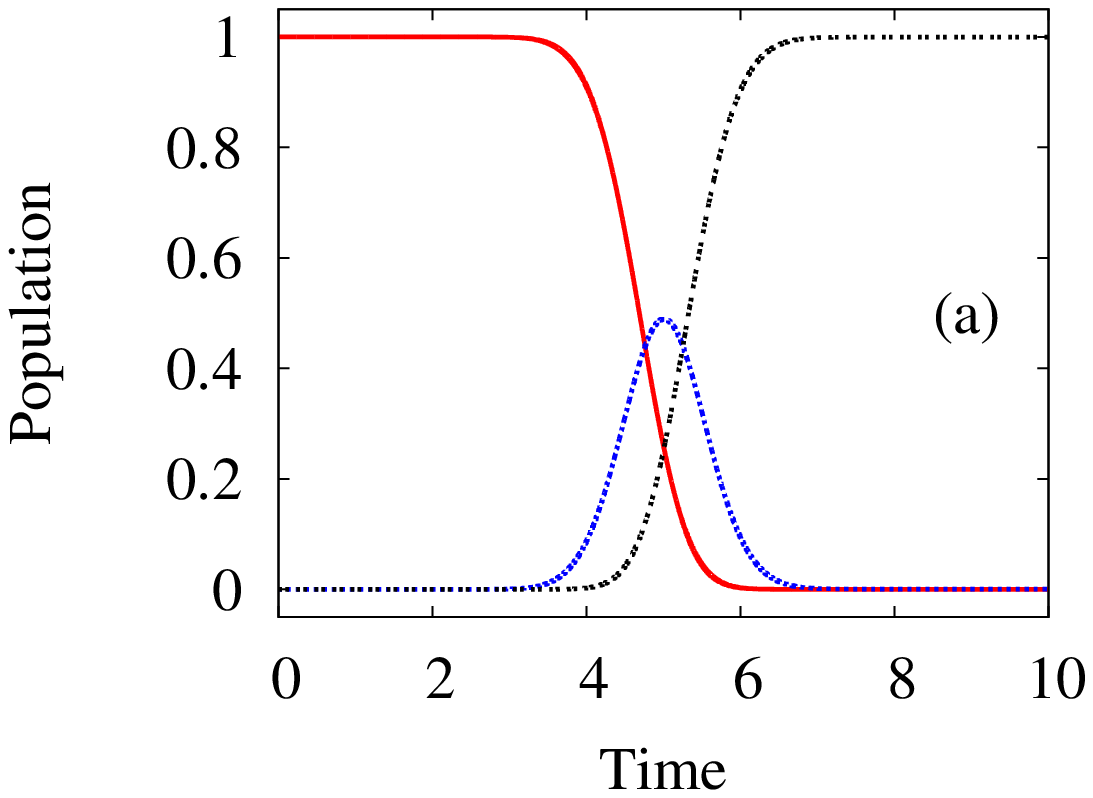} &
\includegraphics[width=0.4\textwidth,angle=0]{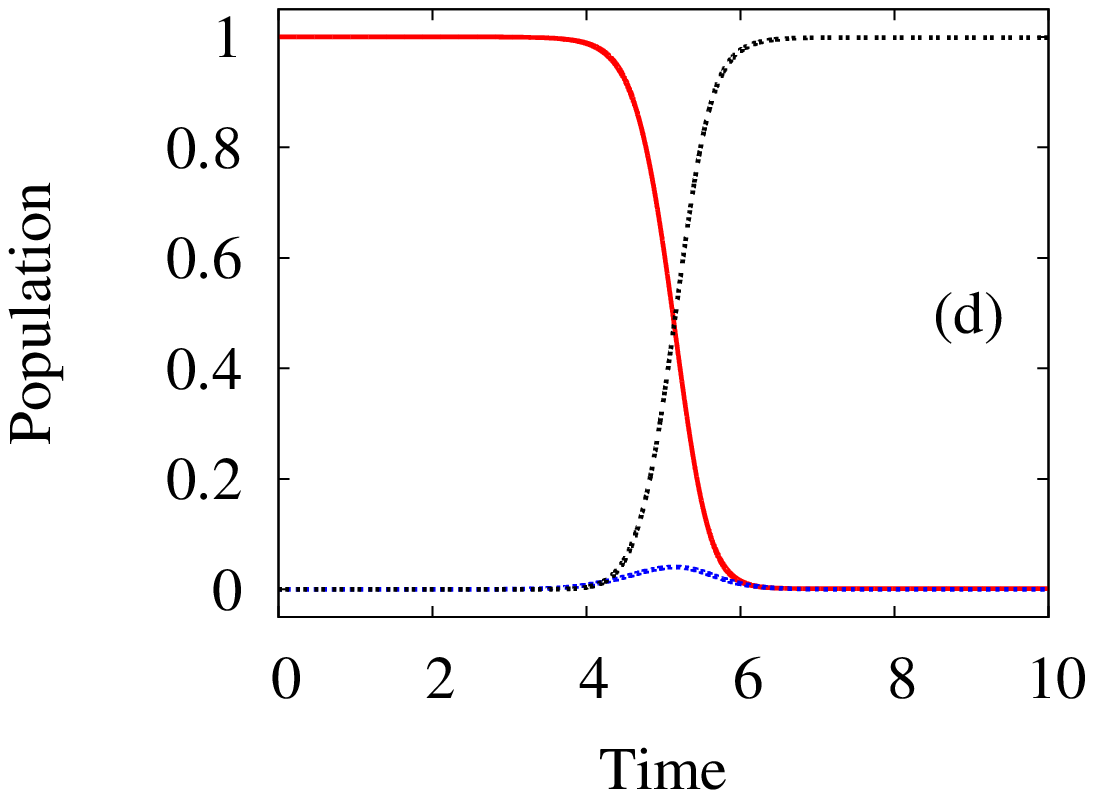} \\
\includegraphics[width=0.4\textwidth,angle=0]{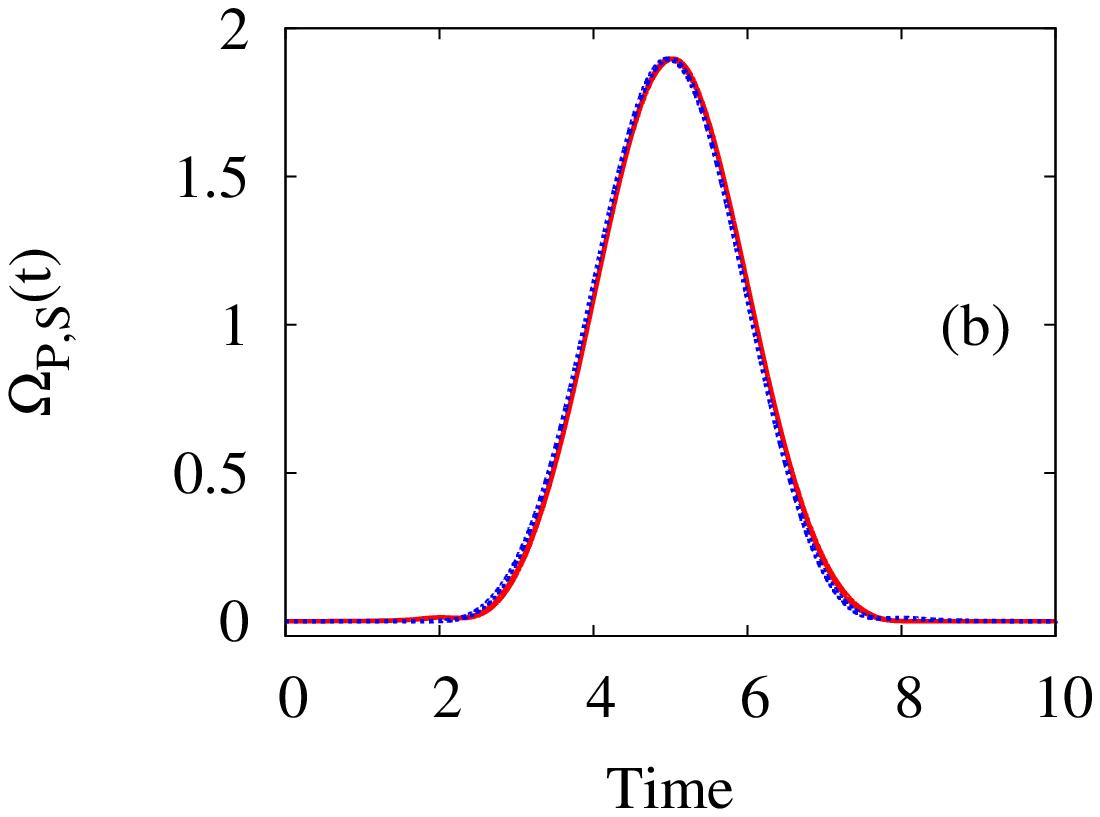} &
\includegraphics[width=0.4\textwidth,angle=0]{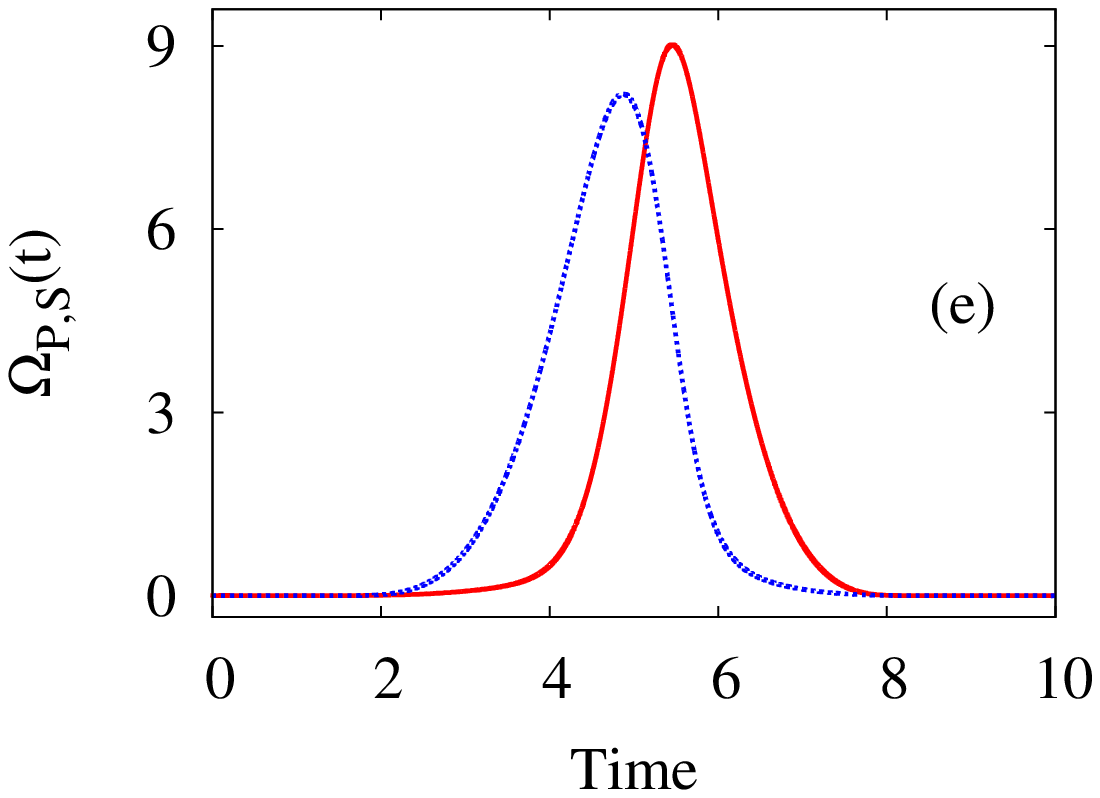} \\
\includegraphics[width=0.4\textwidth,angle=0]{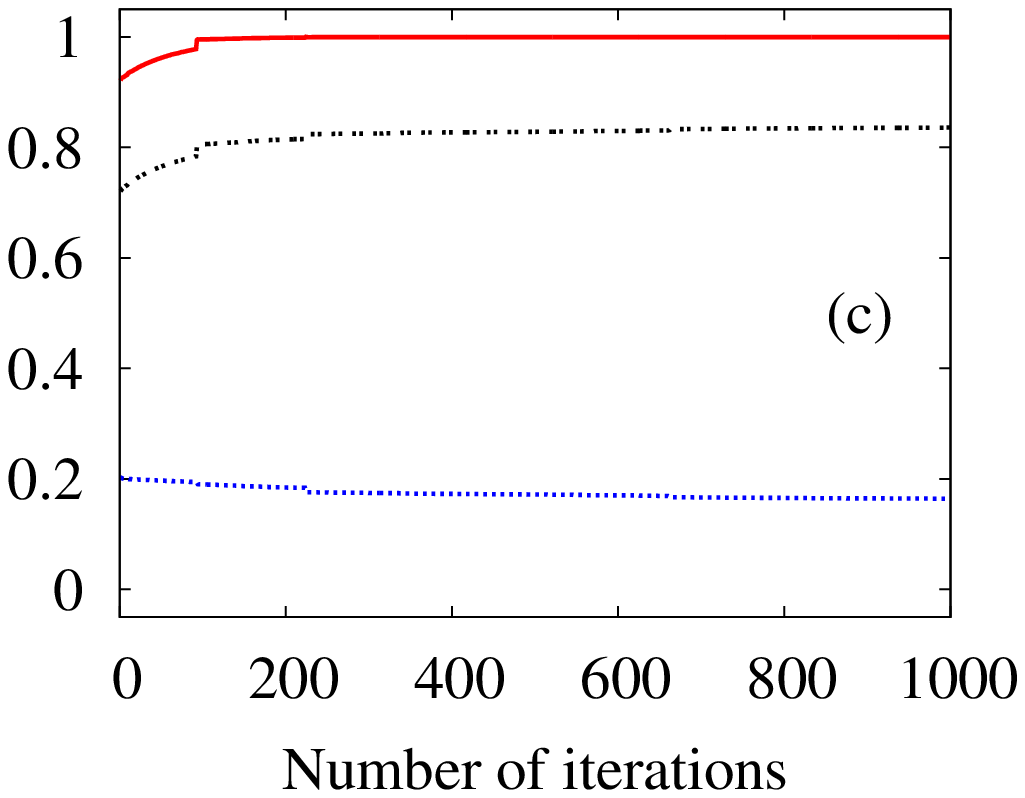} &
\includegraphics[width=0.4\textwidth,angle=0]{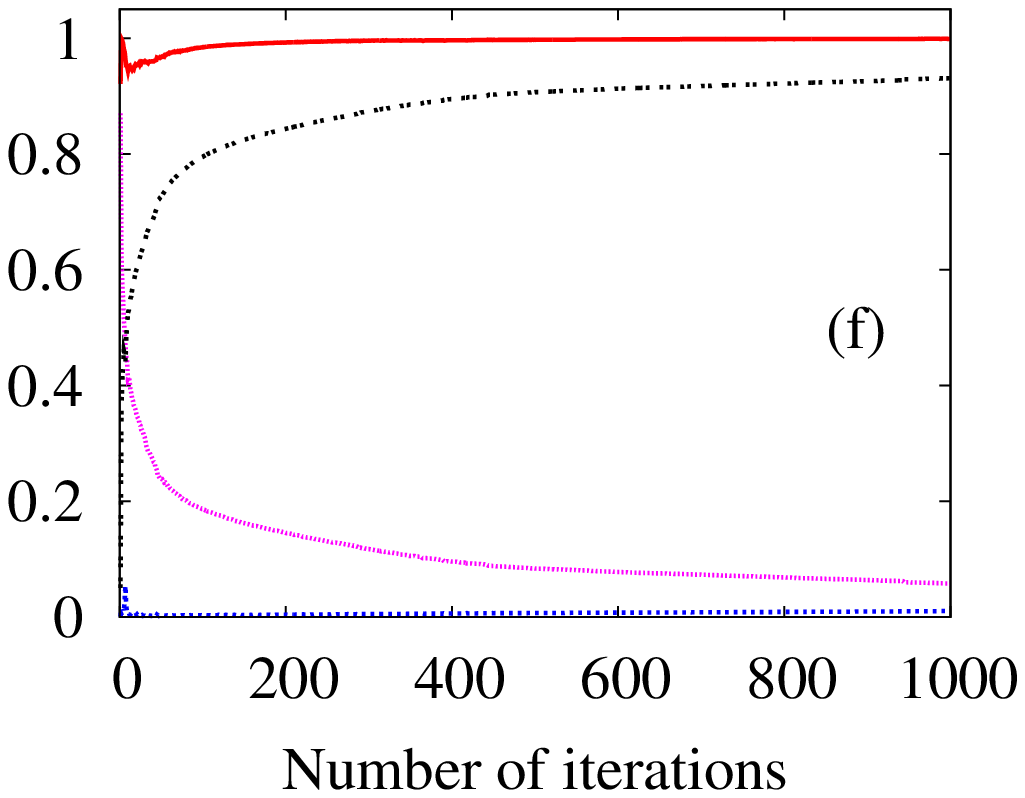} \\
\end{tabular}
%\scriptsize
\caption{ Population transfer in a $\Lambda$ system using conjugate gradient method with penalty on population of the state $|2\rangle$ (right panel) and without penalty, $\beta=0$ (left panel). (a), (d) The population of state $\vert 1 \rangle$ - red line, $\vert 2 \rangle$ - blue line and $\vert 3 \rangle$ - black line. (b), (e) The sequence of optimal pulses: Rabi frequency of the pump pulse - red line and Stokes pulse - blue line. (c), (f) The convergence behavior of the optimized transition probability - red line, penalty on the field energy - blue line, penalty on the intermediate state population - pink line and final optimized cost functional - black line versus the number of iteration steps.}\label{Gradient}
\end{figure}

\begin{figure}[b]
\centering
\begin{tabular}{cc}
\includegraphics[width=0.4\textwidth,angle=0]{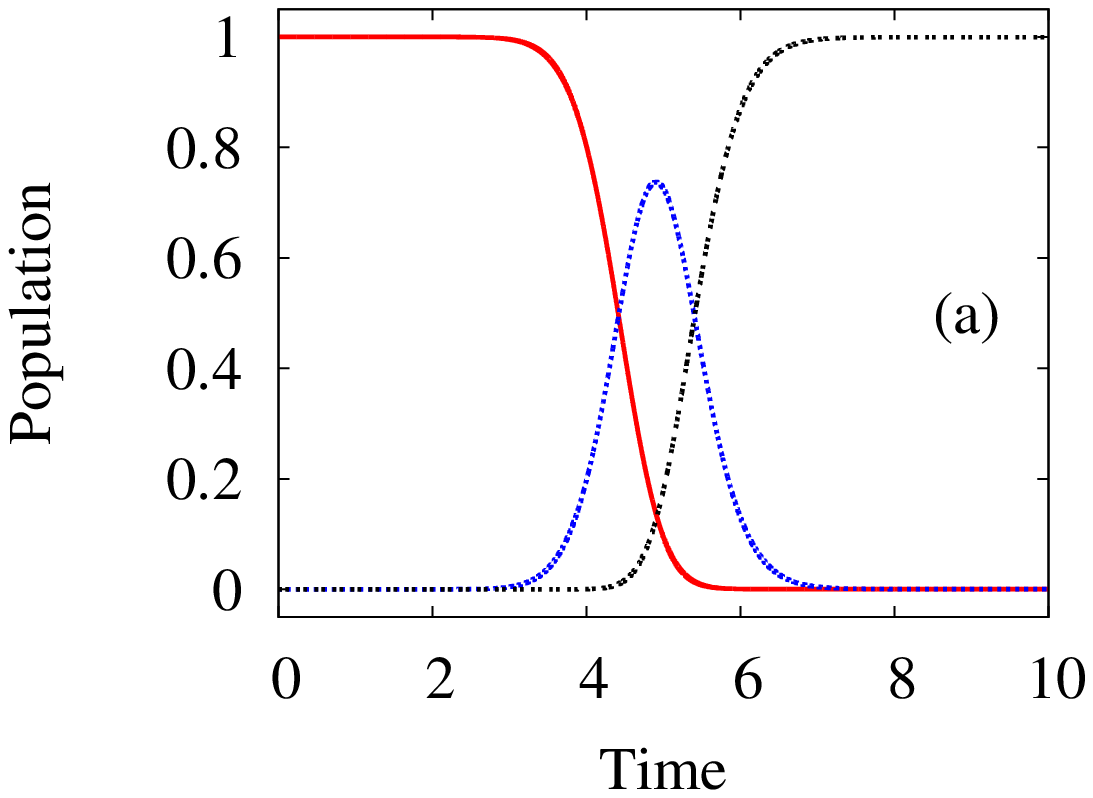} &
\includegraphics[width=0.4\textwidth,angle=0]{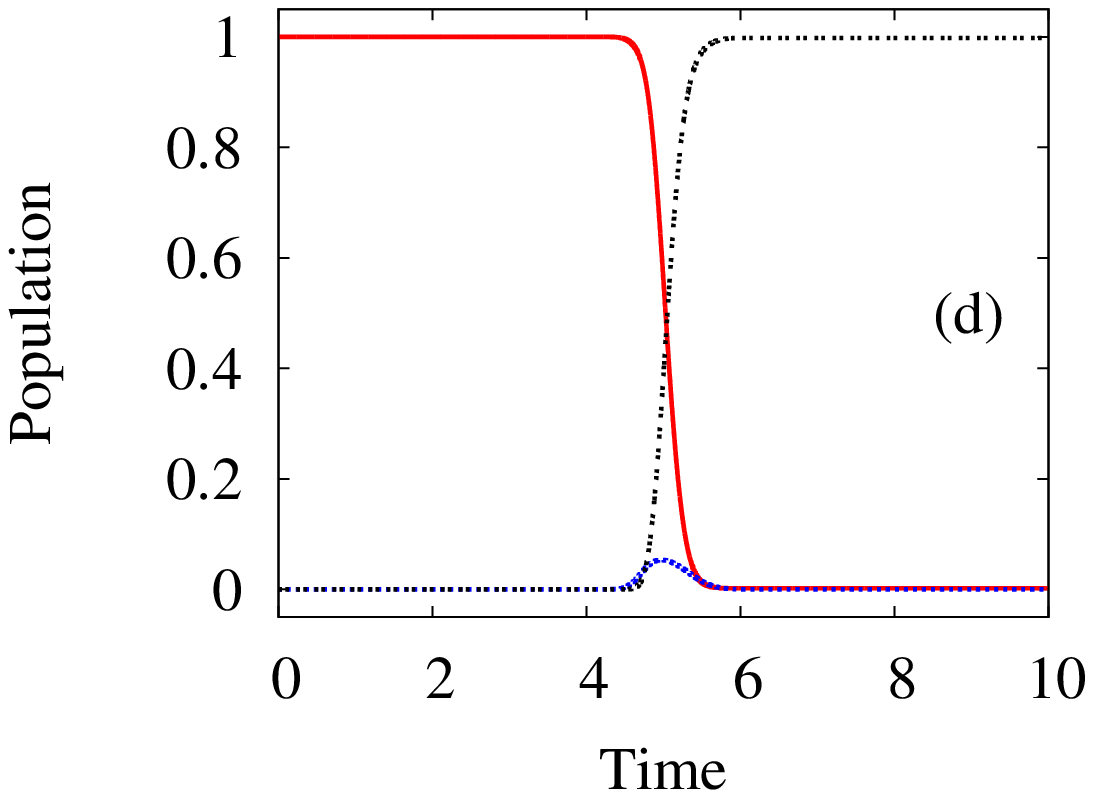} \\
\includegraphics[width=0.4\textwidth,angle=0]{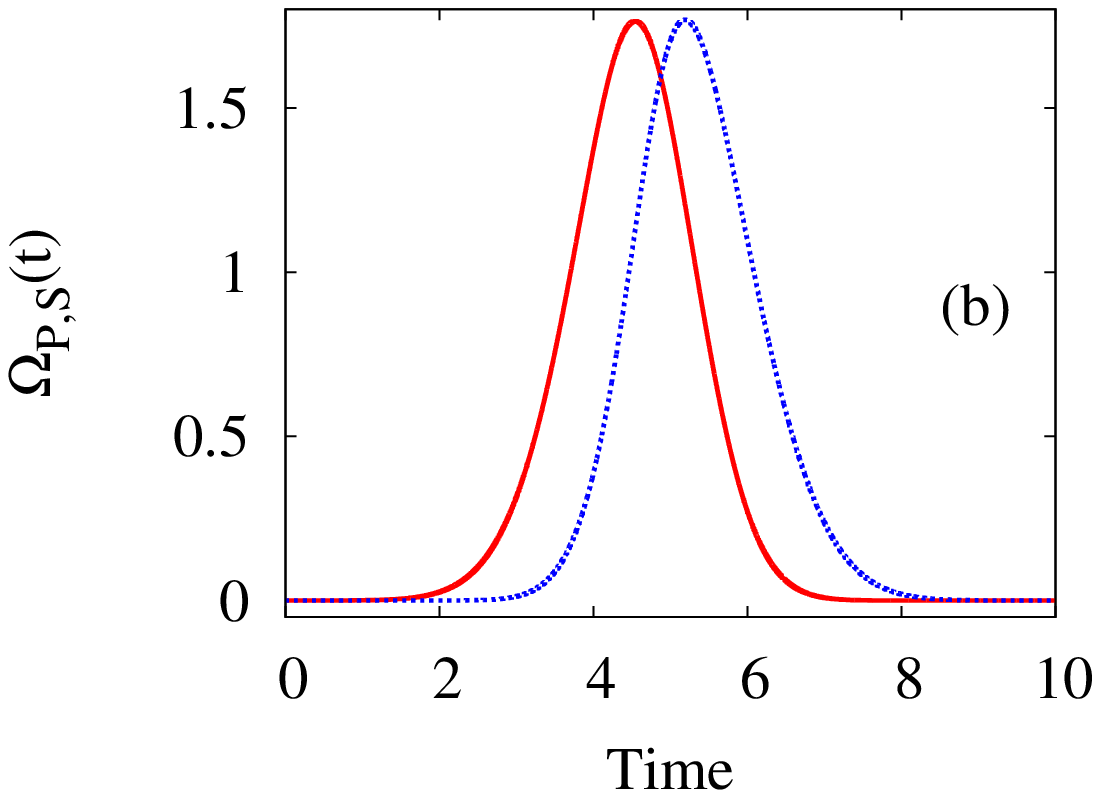} &
\includegraphics[width=0.4\textwidth,angle=0]{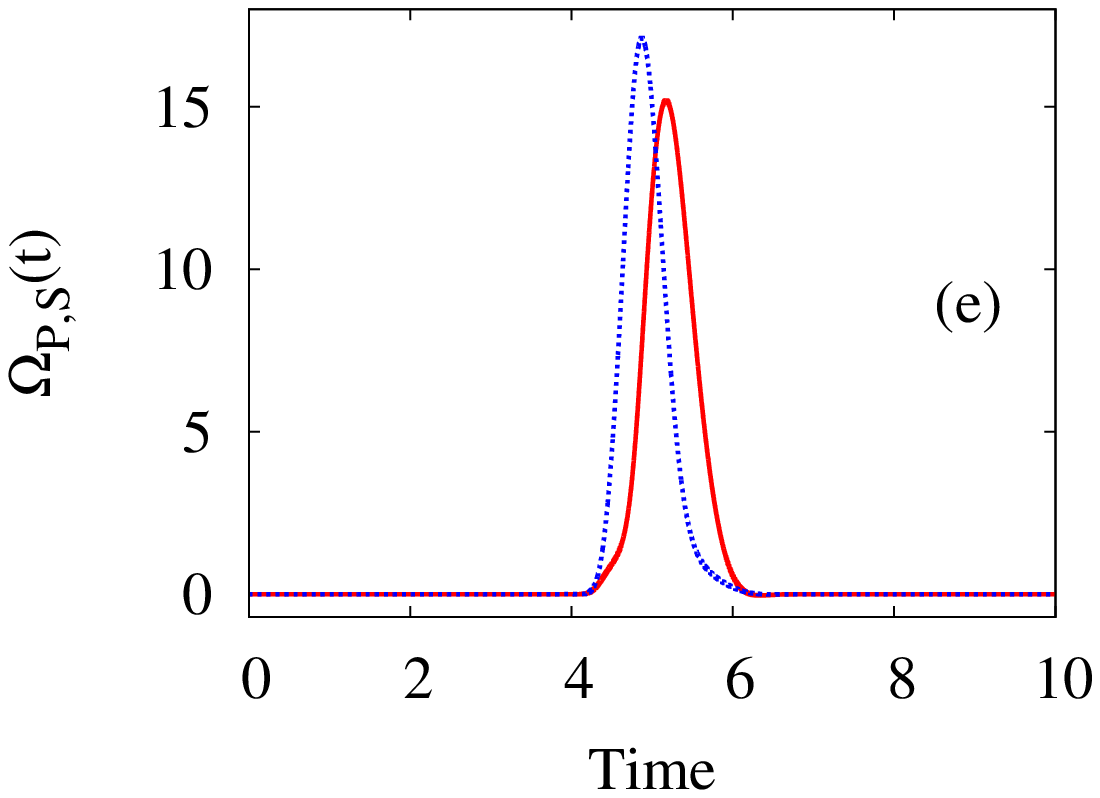} \\
\includegraphics[width=0.4\textwidth,angle=0]{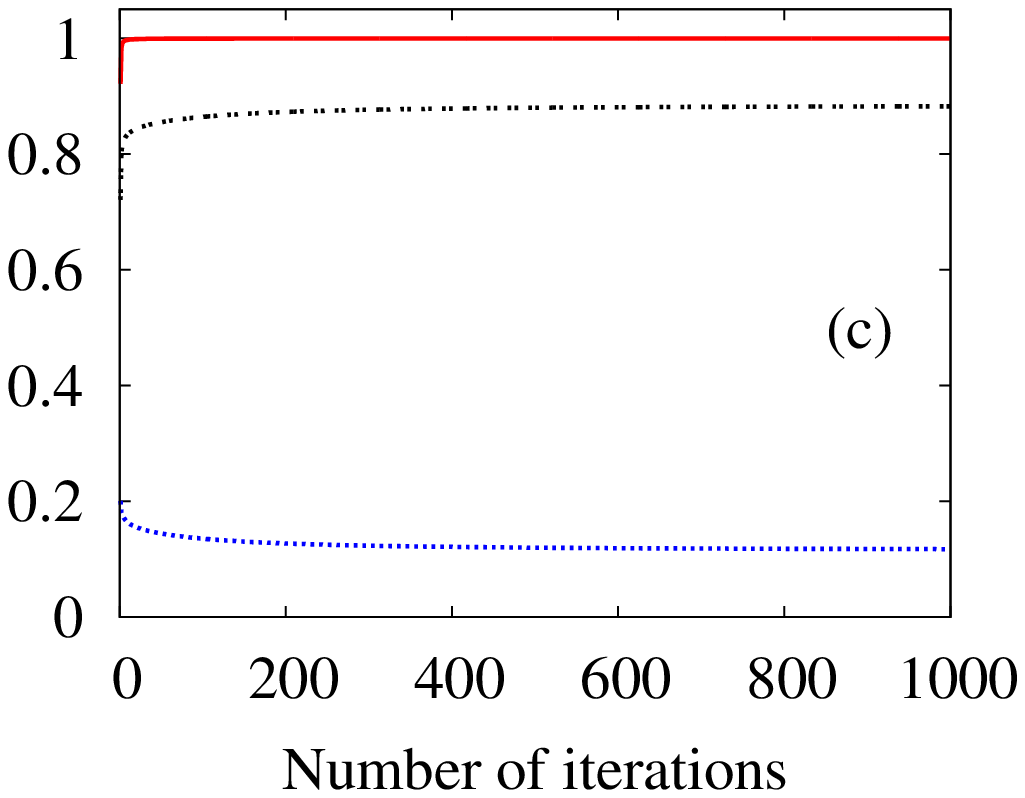} &
\includegraphics[width=0.4\textwidth,angle=0]{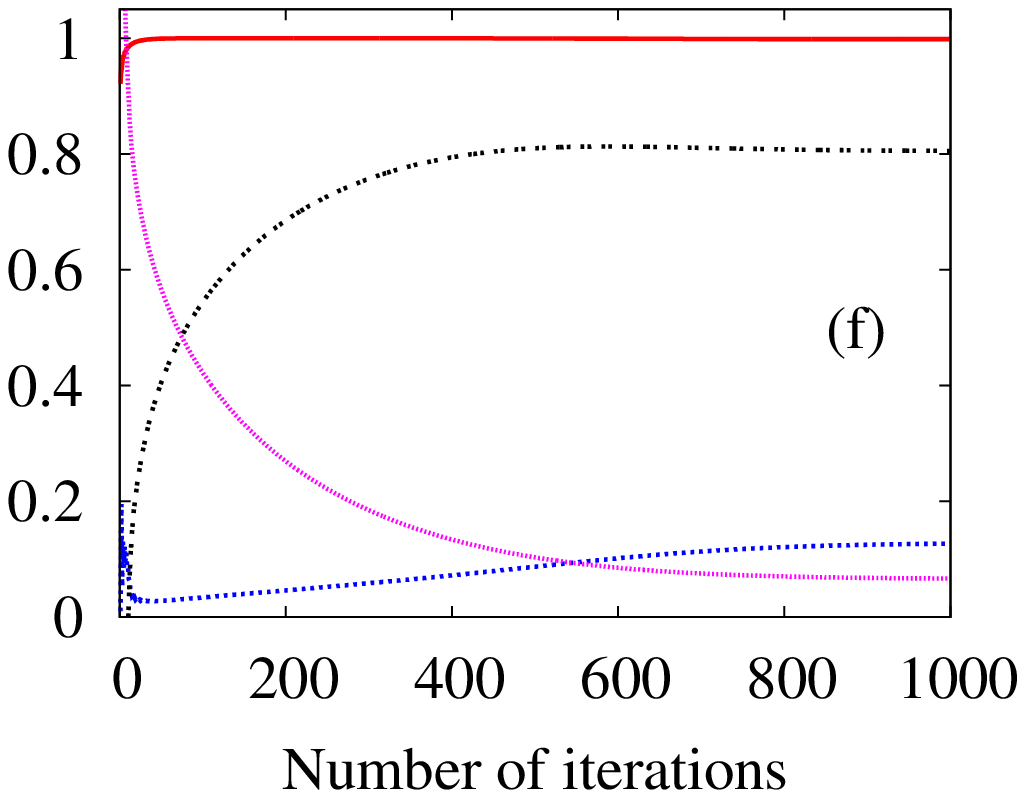} \\
\end{tabular}
\caption{Population transfer in a $\Lambda$ system using Zhu-Rabitz method; description of plots same as in Fig.~\ref{Gradient}.}\label{Rabitz}
\end{figure}
\end{widetext}

{\bf The conjugate gradient method.} The results obtained using the conjugate gradient method~\cite{Kis-2,balint-kurti,press} are shown in Fig.~\ref{Gradient}. The set of plots (a)-(c) shows results obtained without the penalty on the intermediate state population, $\beta = 0$. Plots (d)-(f) show results obtained when the penalty is imposed on the population of state $\vert 2 \rangle$, $\beta \neq 0$. The population of the states $\vert 1 \rangle$, $\vert 2 \rangle$, and $\vert 3 \rangle$ as a function of time is shown in plots (a) and (d). Complete population transfer from initial level $\vert 1 \rangle$ to the target level $\vert 3 \rangle$ is achieved at the final time $T$ independently of the penalty parameter. The amount of population in the intermediate state is considerably reduced for a field obtained with the state-dependent constraint (Fig.~\ref{Gradient}(d)) in contrast to that resulting from unconstrained optimization (Fig.~\ref{Gradient}(a)). The optimized Rabi frequencies (obtained after 1000 iterations) are shown in Fig.~\ref{Gradient} (b) and (e). With a proper choice of the penalty parameters $\alpha_0$ and $\beta$, we remove almost all the intuitive solutions and obtain the STIRAP solution, a counterintuitive pulse sequence when the Stokes pulse precedes the pump. Plots (c) and (f) of Fig.~\ref{Gradient} show the convergence behavior of the optimized transition probability defined as ${\cal P} = \vert \langle \psi_{}(T) \vert \phi_{}(T) \rangle \vert^2$, and the cost functional, $K$, versus the number of iteration steps. It is seen that the transition probability reaches nearly 100\%.

\begin{widetext}
\begin{figure}[b]
\centering
\begin{tabular}{cc}
\includegraphics[width=0.4\textwidth,angle=0]{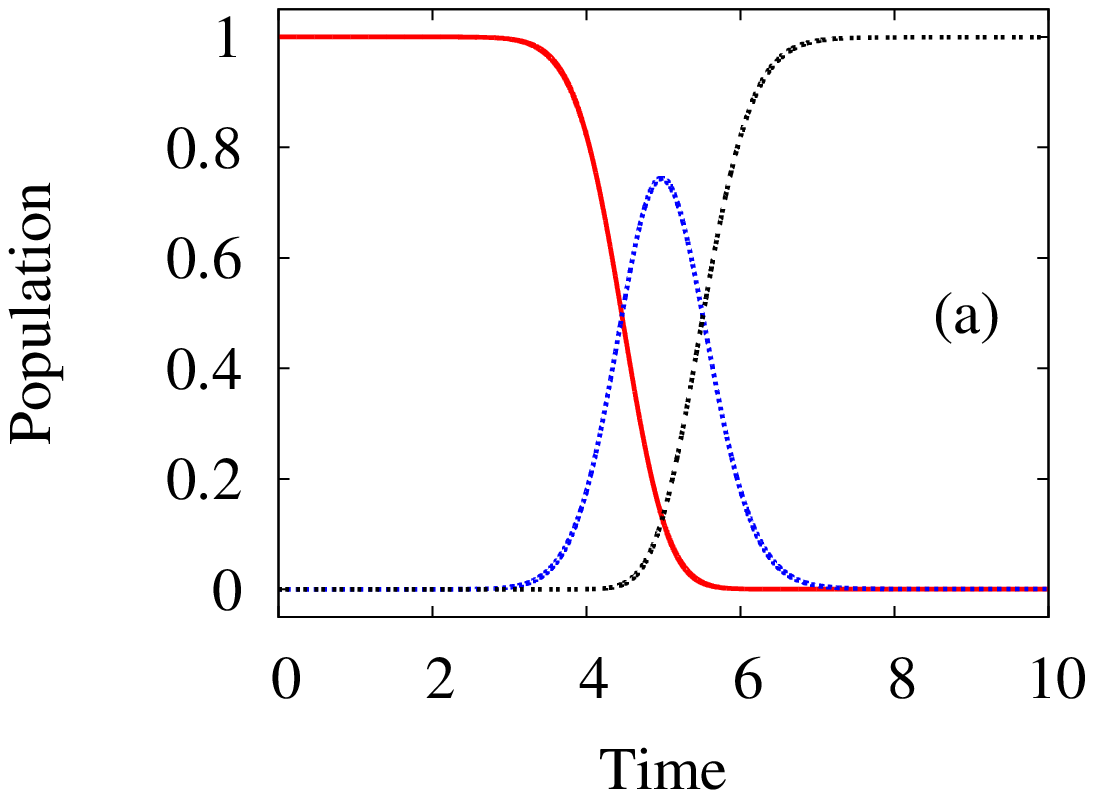} &
\includegraphics[width=0.4\textwidth,angle=0]{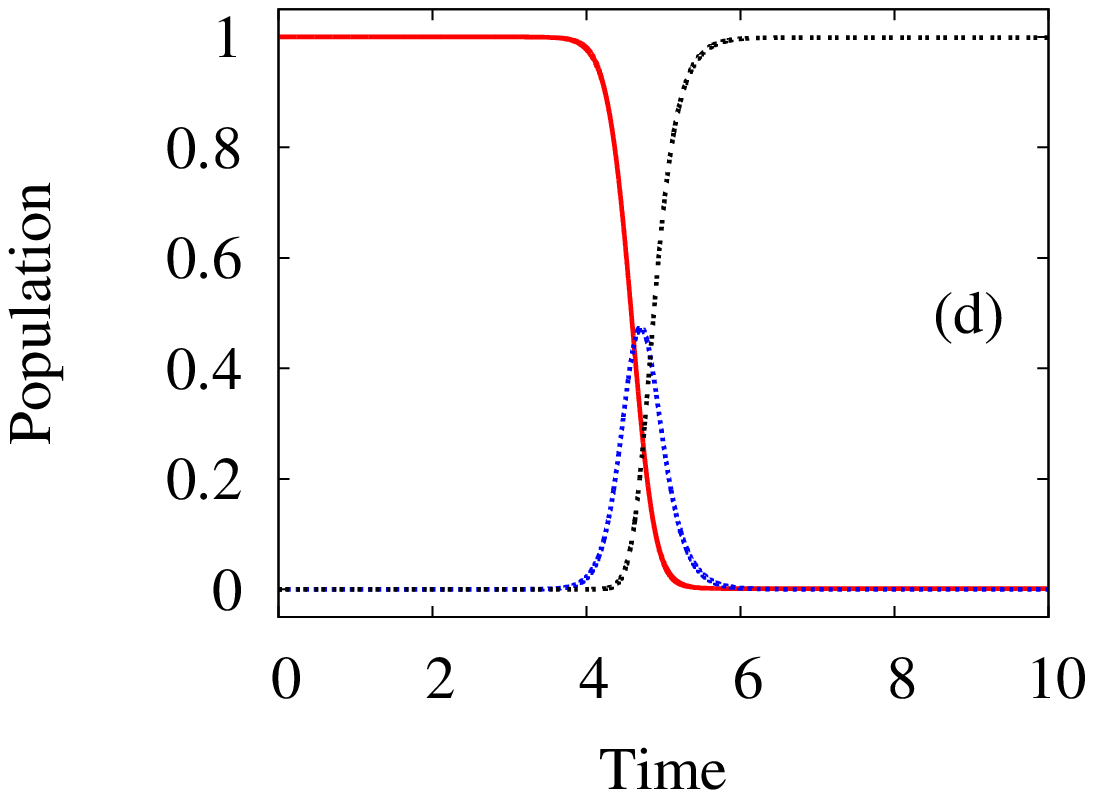}  \\
\includegraphics[width=0.4\textwidth,angle=0]{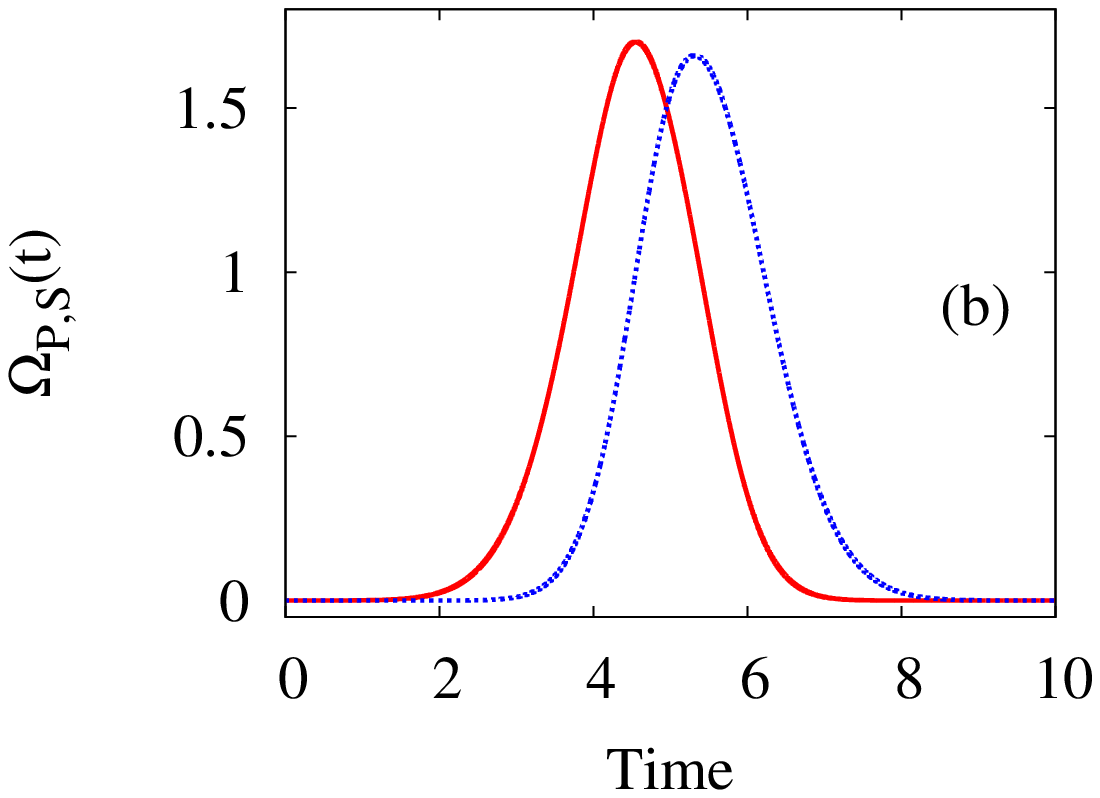} &
\includegraphics[width=0.4\textwidth,angle=0]{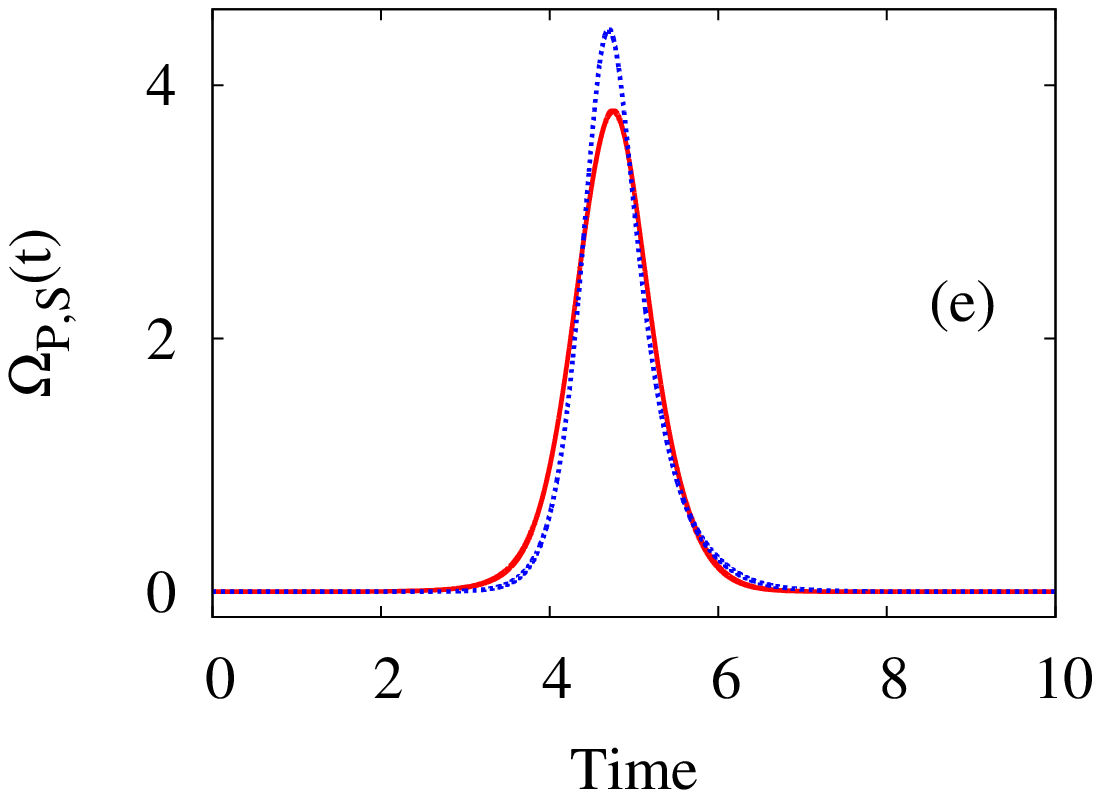}   \\
\includegraphics[width=0.4\textwidth,angle=0]{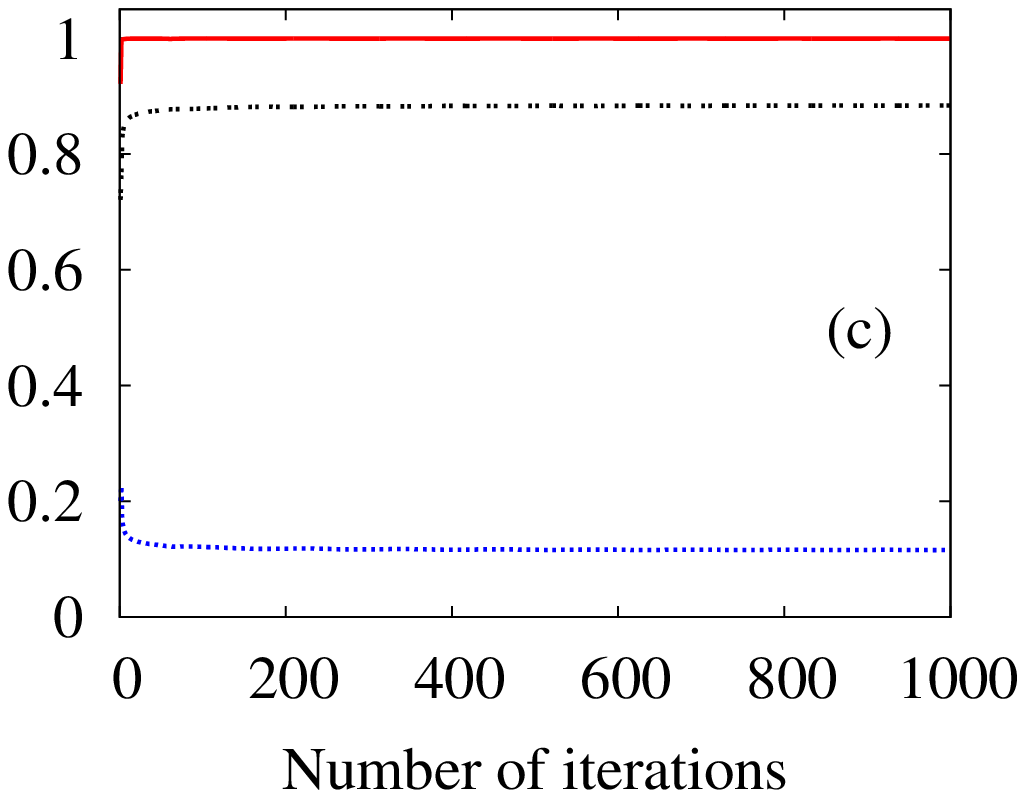} &
\includegraphics[width=0.4\textwidth,angle=0]{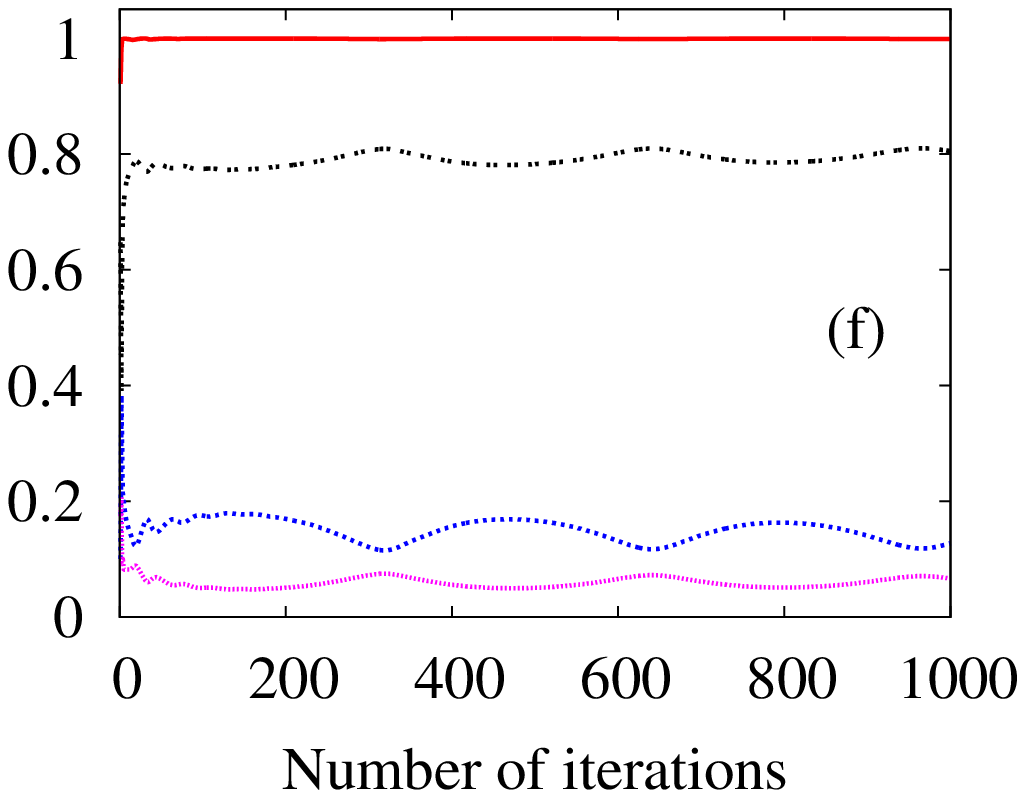}  \\
\end{tabular}
\caption{Population transfer in a $\Lambda$ system using Krotov method:
description of plots same as in Fig.~\ref{Gradient}.}\label{Krotov}
\end{figure}
\end{widetext}

\begin{widetext}
\begin{figure}[b]
\centering
\begin{tabular}{cc}
\includegraphics[width=0.4\textwidth,angle=0]{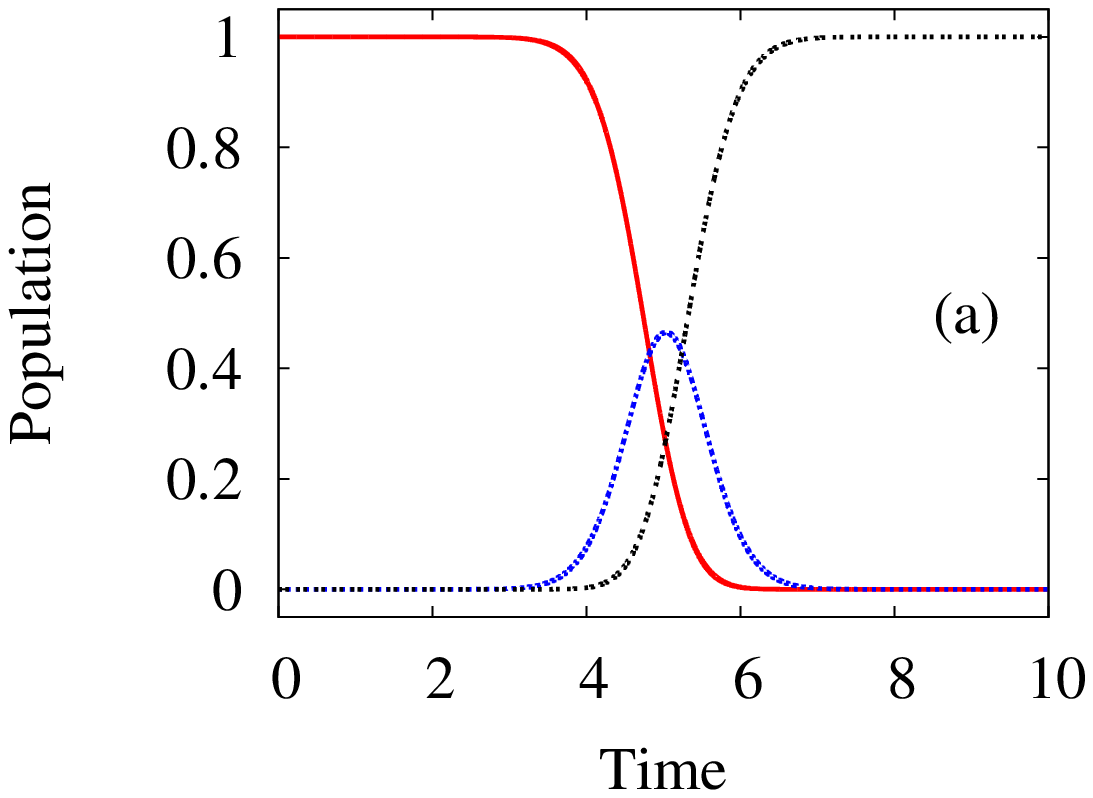} &
\includegraphics[width=0.4\textwidth,angle=0]{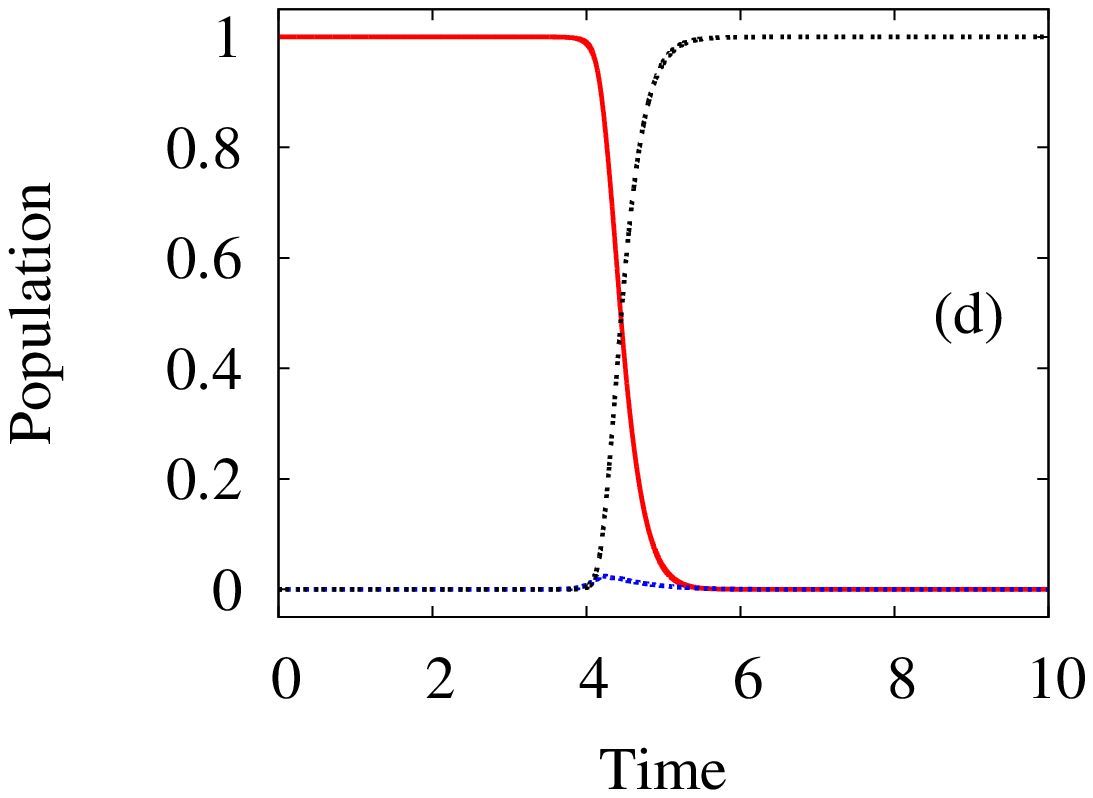} \\
\includegraphics[width=0.4\textwidth,angle=0]{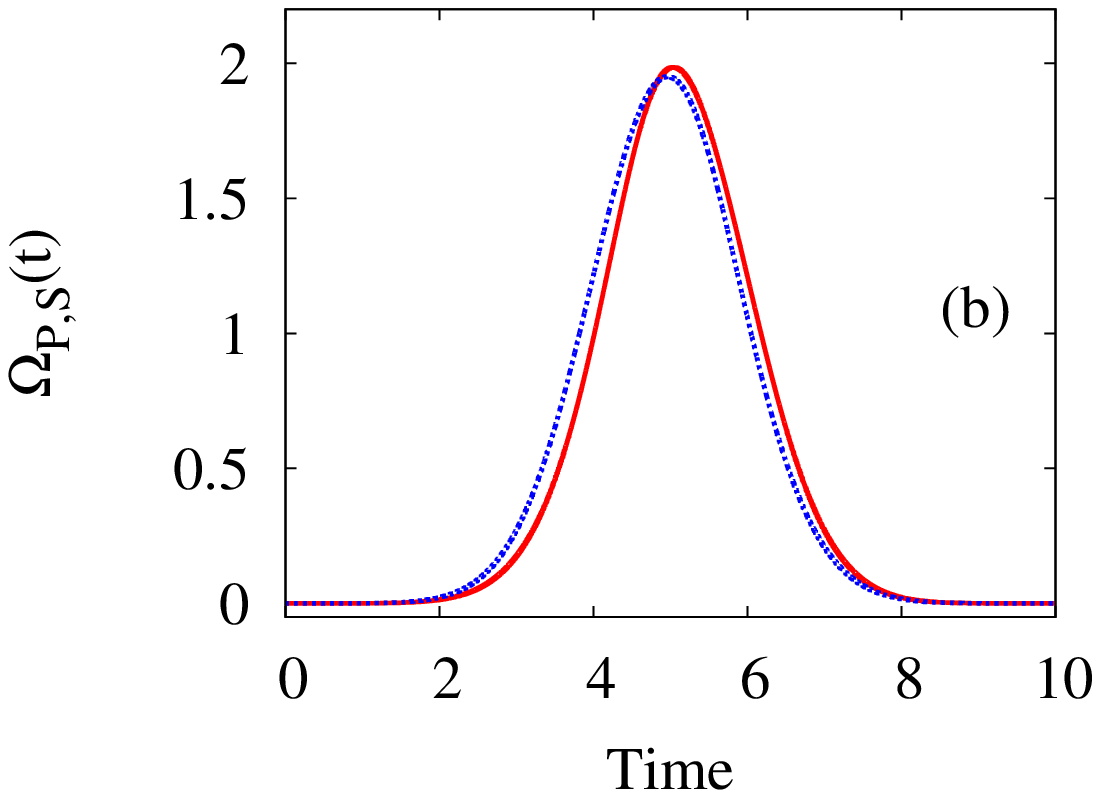} &
\includegraphics[width=0.4\textwidth,angle=0]{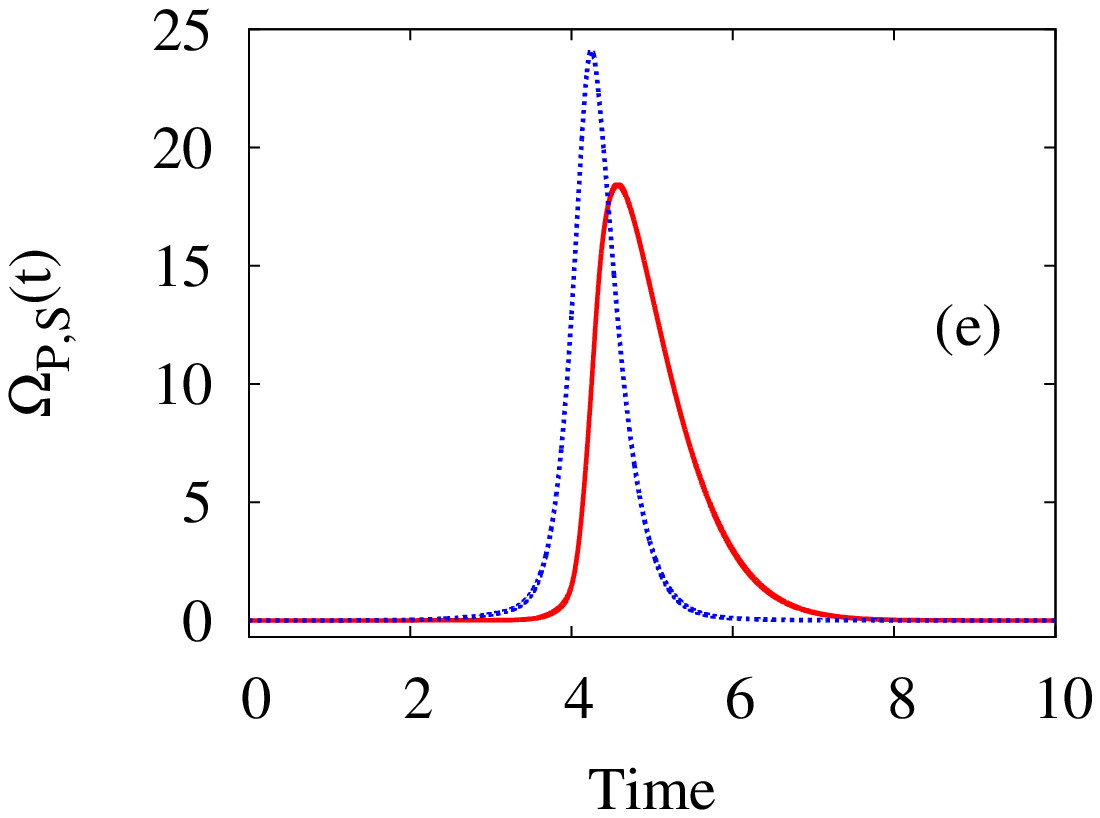} \\
\includegraphics[width=0.4\textwidth,angle=0]{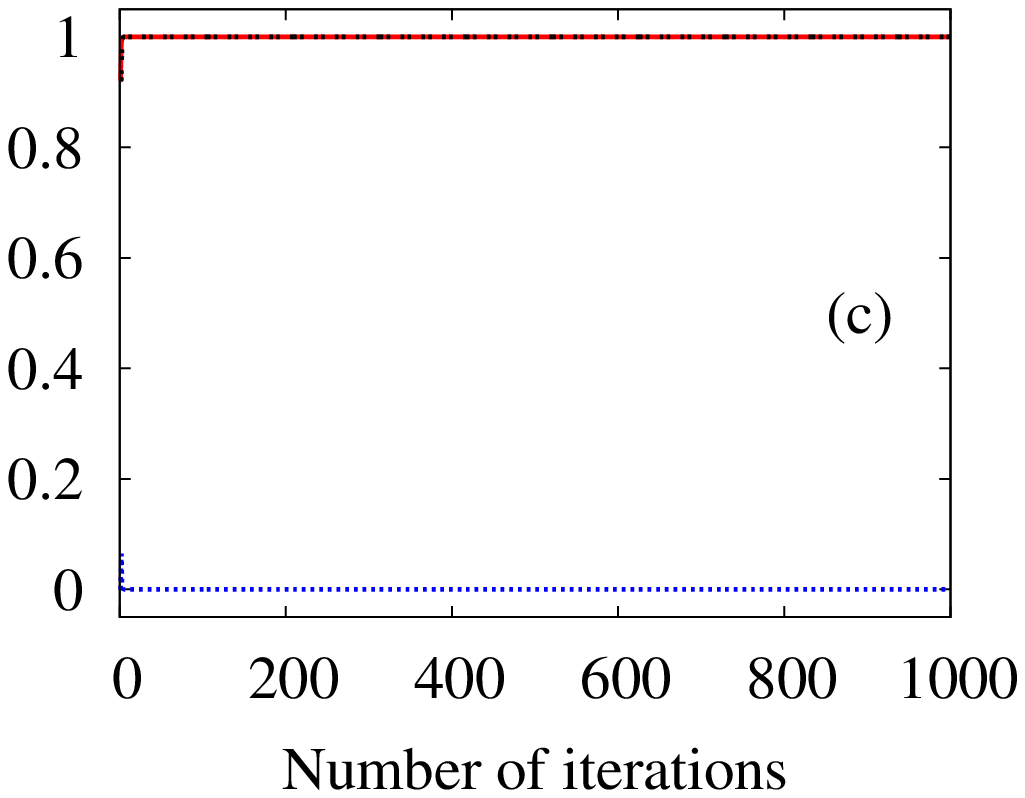} &
\includegraphics[width=0.4\textwidth,angle=0]{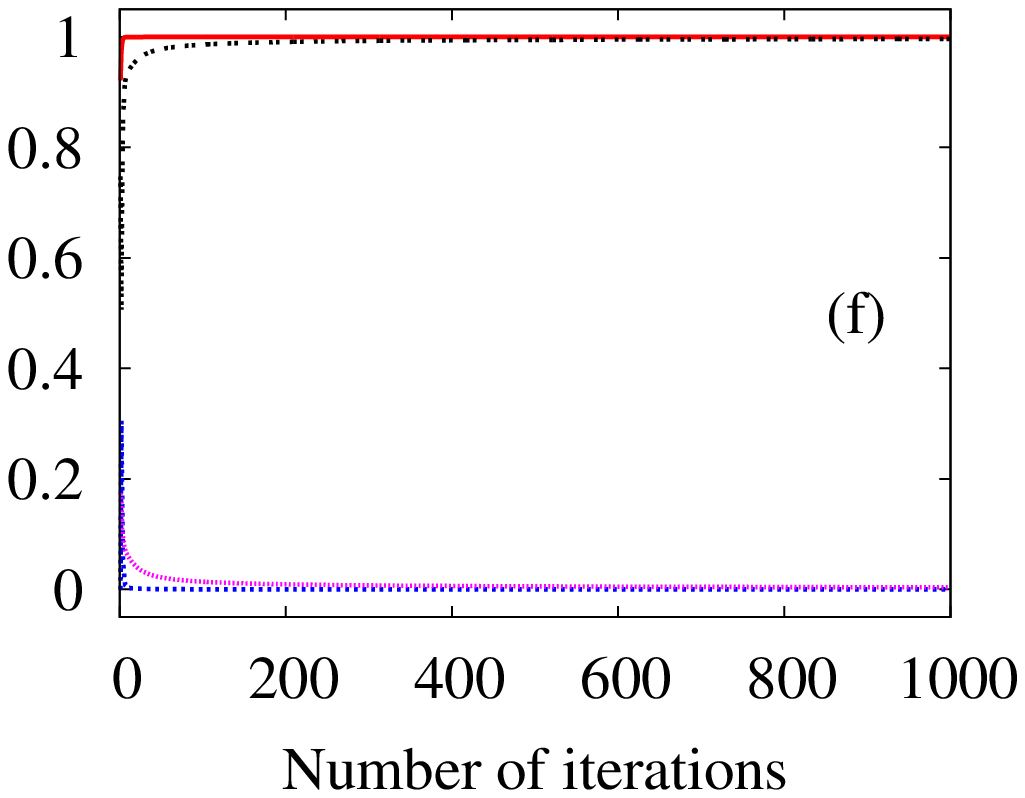} \\
\end{tabular}
\caption{Population transfer in a $\Lambda$ system using the Krotov method for $\Omega_{P,S}^r(t) \neq 0$; description of plots same as in Fig.~\ref{Gradient}.}\label{ronnie}
\end{figure}
\end{widetext}
{\bf The Zhu-Rabitz method.}
Figure~\ref{Rabitz} shows the dynamics of population transfer in the three-level $\Lambda$ system optimized by using the Zhu-Rabitz method~\cite{zhu}. The same order of illustrations and legends is kept in Fig.~\ref{Rabitz} as in the case of Fig.~\ref{Gradient}.

There is a complete transfer of the population from the initial state to the target state at final time $T$, Fig.~\ref{Rabitz} (a) and (d).
However, we observe two different mechanisms of population transfer. Optimization without a penalty on the intermediate state population (left panel of Fig.~\ref{Rabitz}) produces an intuitive pulse sequence, first the pump, then the Stokes pulse is applied to the system (Fig.~\ref{Rabitz}(b)). That pulse arrangement results in a sequential population transfer from the state $|1\rangle$ to the state $|2\rangle$ and then to the target state $|3\rangle$ as it is demonstrated in Fig.~\ref{Rabitz}(a), with almost 80\% of population resided in the state $|2\rangle$ at the central time.

Applying a penalty on the second state population changes the mechanism dramatically (see right panel of Fig.~\ref{Rabitz}). Now we obtained almost hundred times more intense pulses, compare Fig.~\ref{Rabitz} (b) and (e). More importantly,
we automatically obtain a counterintuitive pulse sequence when the Stokes pulse precedes the pump pulse. Therefore, it is clear that the OCT algorithm finds a well known STIRAP solution which has a highly pronounced signature of suppressed intermediate state population, see Fig.~\ref{Rabitz} (d). In that case population transfer takes place through the dark state consisting ideally only of the initial and the target state probability amplitudes and has no projection to the state $|2\rangle$~\cite{bergmann}.

The value of the optimized cost functional in Fig.~\ref{Rabitz} (f) is a bit less than that of the cost functional value obtained by using the conjugate gradient method, Fig.~\ref{Gradient} (f). This is due to a slightly larger energy fluency of the fields as it is clear from the field amplitudes in the plot (e) in Fig.~\ref{Rabitz} comparing to that in Fig.~\ref{Gradient}.

{\bf The Krotov method.} The optimized results of population transfer using the Krotov method~\cite{somloi,tannor} are shown in Figs.~\ref{Krotov} and~\ref{ronnie}. The same set of parameters is used in the optimization process as in the previous two cases. Figure~\ref{ronnie} illustrates the results obtained using the Krotov method when $\Omega_{P,S}^{r}(t) \neq 0$~\cite{palao}. Population of the states, optimized Rabi frequencies, transition probability and cost functional are shown in plots (a)-(c) and (d)-(f) of Figs.~\ref{Krotov} and~\ref{ronnie}.

As in the previous cases, we produce complete population transfer to the target state. However, using Krotov method~\cite{somloi,tannor} with $\Omega_{P,S}^{r}(t) = 0$ we were not able to find a set of penalty parameters which provides considerable suppression of the second state population, see Fig.~\ref{Krotov}~(d). At most we are able to reduce the second state population to only about 40\%. Optimized Rabi frequencies are arranged in the intuitive manner even for the reasonably large areas of the pulses, Fig.~\ref{Krotov}~(b) and (e).

The situation changes considerably when we use the reference fields,  $\Omega_{P,S}^{r}(t) \neq 0$, Fig.~\ref{ronnie}. In the manner similar to previous two cases the STIRAP solution emerges from the optimization procedure when we impose a penalty on the intermediate state population, see right panel in Fig~\ref{ronnie}: the Stokes pulse precedes the pump Fig~\ref{ronnie}~(e) and a detrimental population of the state $|2\rangle$ is suppressed almost to zero, Fig~\ref{ronnie}~(d).

Table~\ref{tab:table1} shows a comparison of transition probability ${\cal P}$, optimized cost functional $K$ and the maximum population, $\varrho_{22}=|a_{2}|^{2}$, which resides in the intermediate state at central times for all the methods. Note that the value of the cost functional obtained using the Krotov method for $\Omega_{P,S}^{r}(t) \neq 0$ is larger than corresponding value in other methods. However, the computation cost of the conjugate gradient method is larger than that of the Zhu-Rabitz method and the Krotov method.

\begin{widetext}
\begin{table}[h]
\begin{center}
\caption{\label{tab:table1} Comparison of results obtained using different implementations of the OCT for the population transfer dynamics in the three-level $\Lambda$ system.}
\vskip 5 pt
\begin{tabular}{|c|c|c|c|c|c|c|c|c|c|c|}
\hline
{\it Method} & \multicolumn{5}{c|}{ Without state-dependent penalty} &  \multicolumn{5}{c|}{ With state-dependent penalty } \\ \cline{2-11}
 & $\alpha_0$ & $\beta$ & ${\cal P}$ & $K$ & $\varrho_{22}$ & $\alpha_0$ & $\beta$ & $P$ & $K$ & $\varrho_{22}$ \\
\hline
{\bf Conjugate gradient}           &  0.01  &  0  &  0.999  &  0.836  &  0.48  &  0.00005  &  1.0    &  0.998  &  0.930 &  0.04 \\
\hline
{\bf Zhu-Rabitz}                   &  0.01  &  0  &  0.999  &  0.882  &  0.74  &  0.0005   &  1.8  &  0.998  &  0.805   &  0.053  \\
\hline
{\bf Krotov } ($\Omega_{P,S}^{r}(t) = 0$)  &  0.01  &  0  &  0.999  &  0.883  &  0.74  &  0.005    &  0.2  &  0.998  &  0.805   &  0.47  \\
\hline
{\bf Krotov } ($\Omega_{P,S}^{r}(t) \neq 0$) &  1.0   &  0  &  0.999  &  0.999  &  0.46  &  0.05     &  0.2  &  0.999  &  0.996  &  0.024 \\
\hline
\end{tabular}
\end{center}
\end{table}
\end{widetext}

\subsection{Optimal control to maximize coherence}

\begin{figure}[hbt]
\centering
\begin{tabular}{ccc}
\includegraphics[width=0.285\textwidth,angle=0]{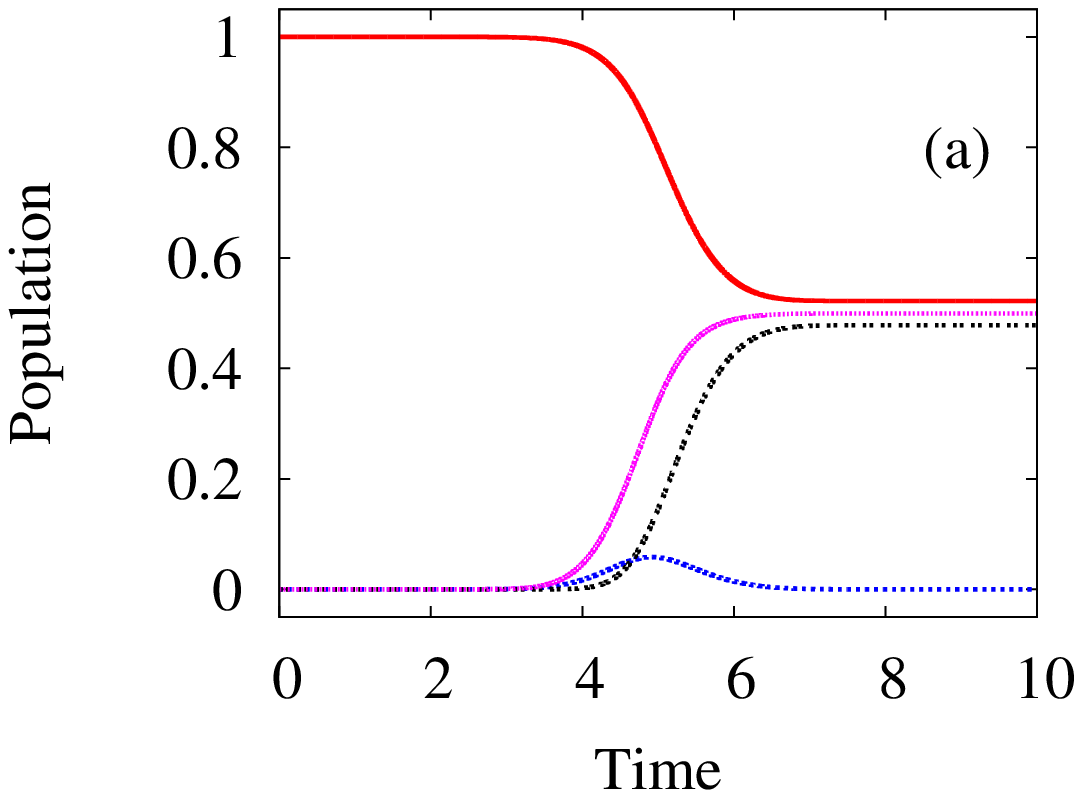} &
\includegraphics[width=0.285\textwidth,angle=0]{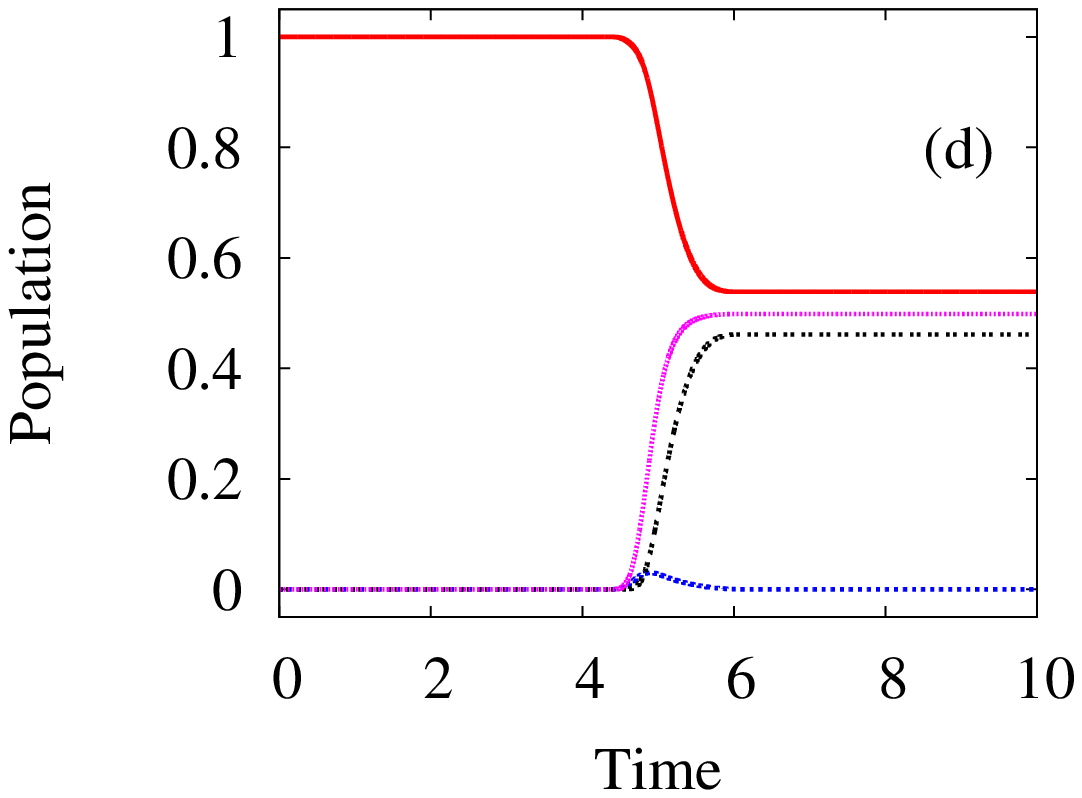} &
\includegraphics[width=0.285\textwidth,angle=0]{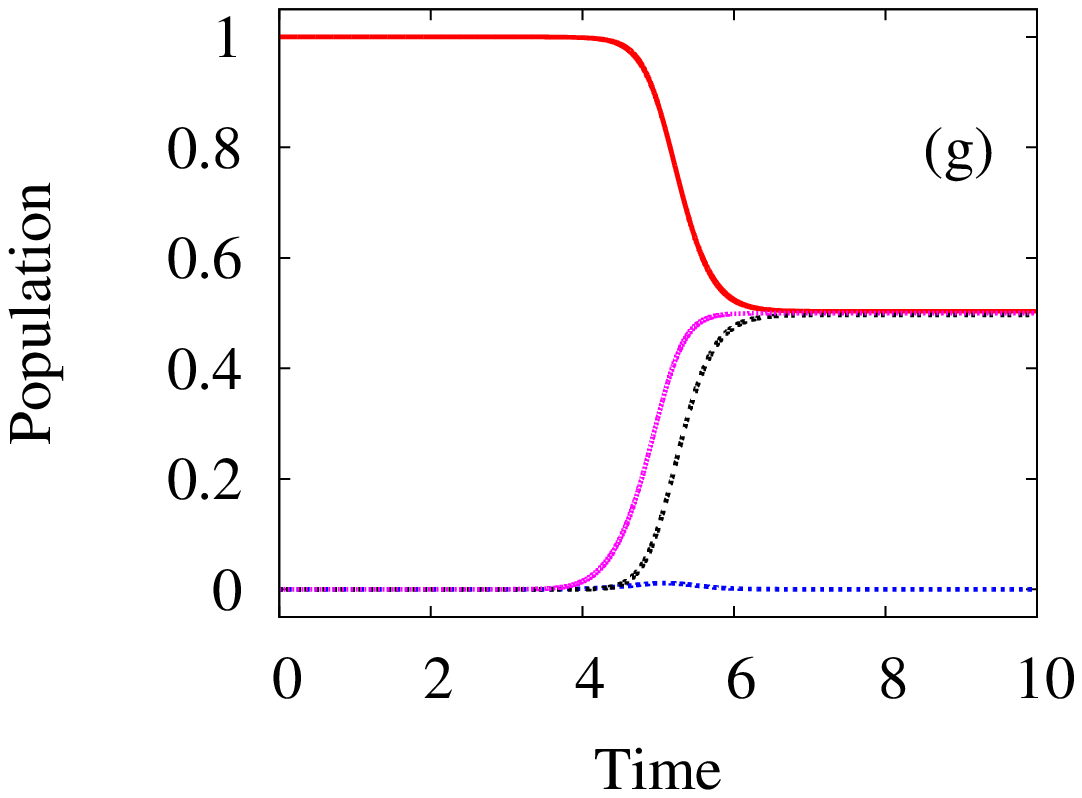} \\
\includegraphics[width=0.285\textwidth,angle=0]{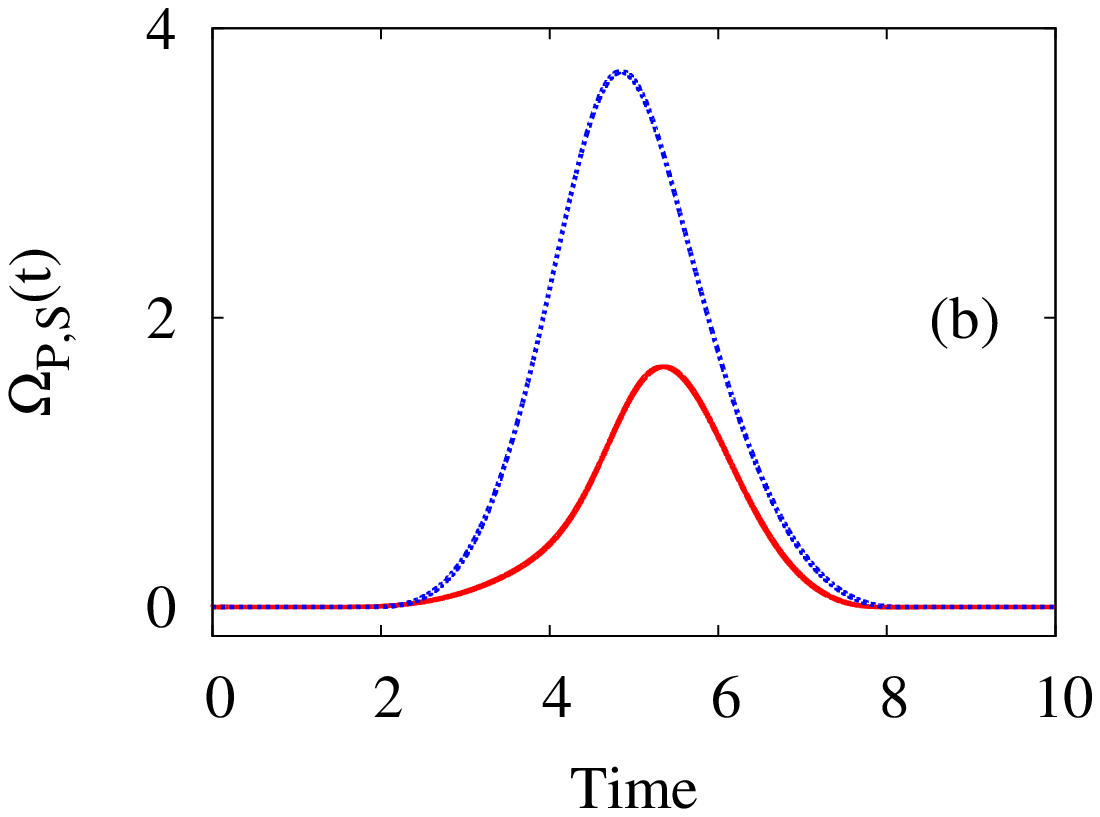} &
\includegraphics[width=0.285\textwidth,angle=0]{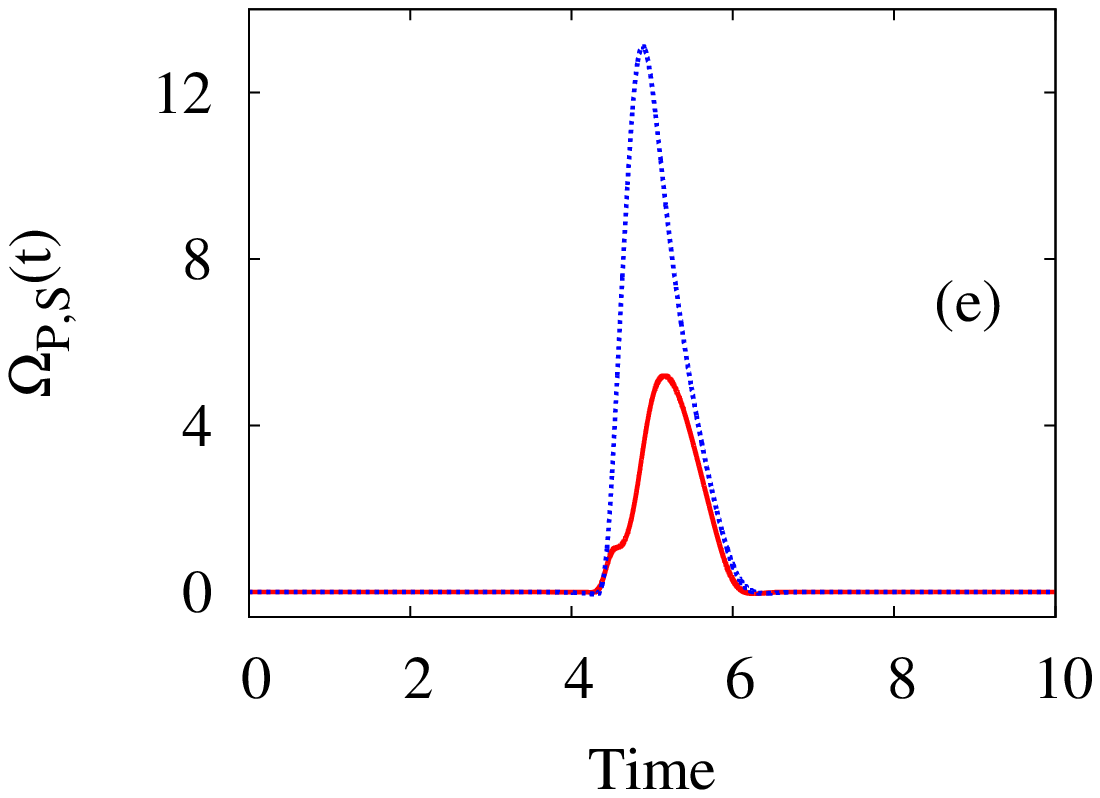} &
\includegraphics[width=0.285\textwidth,angle=0]{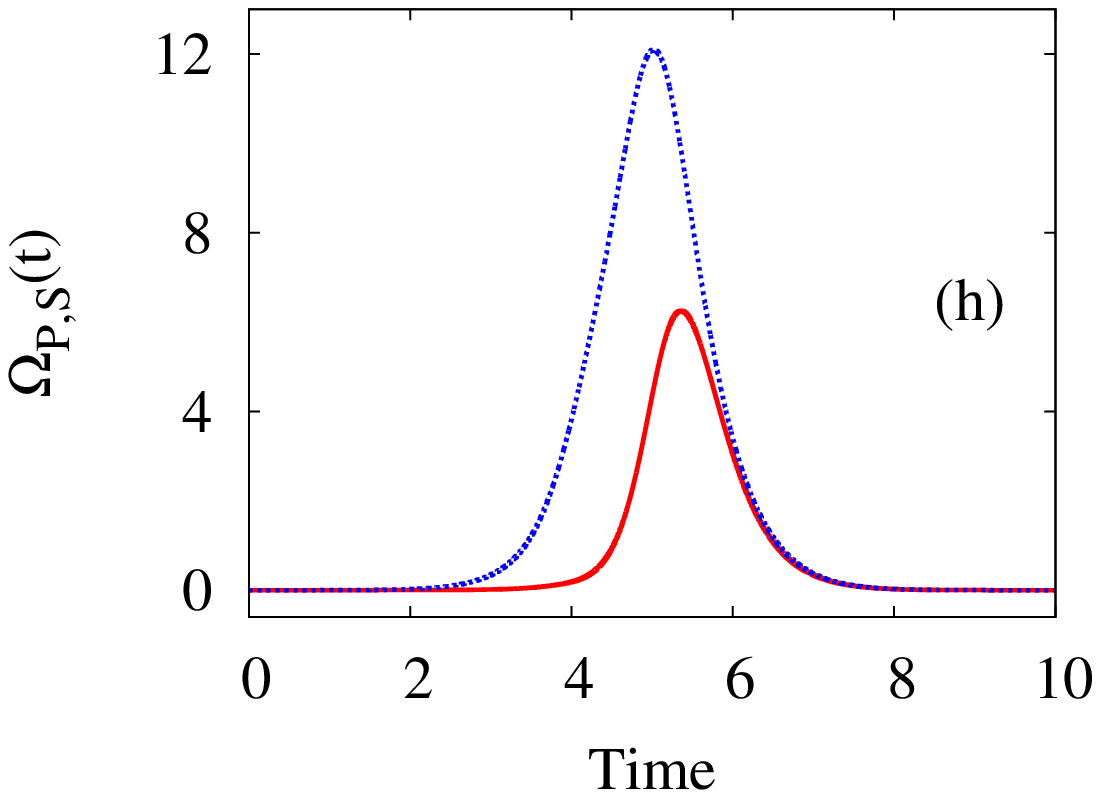} \\
\includegraphics[width=0.285\textwidth,angle=0]{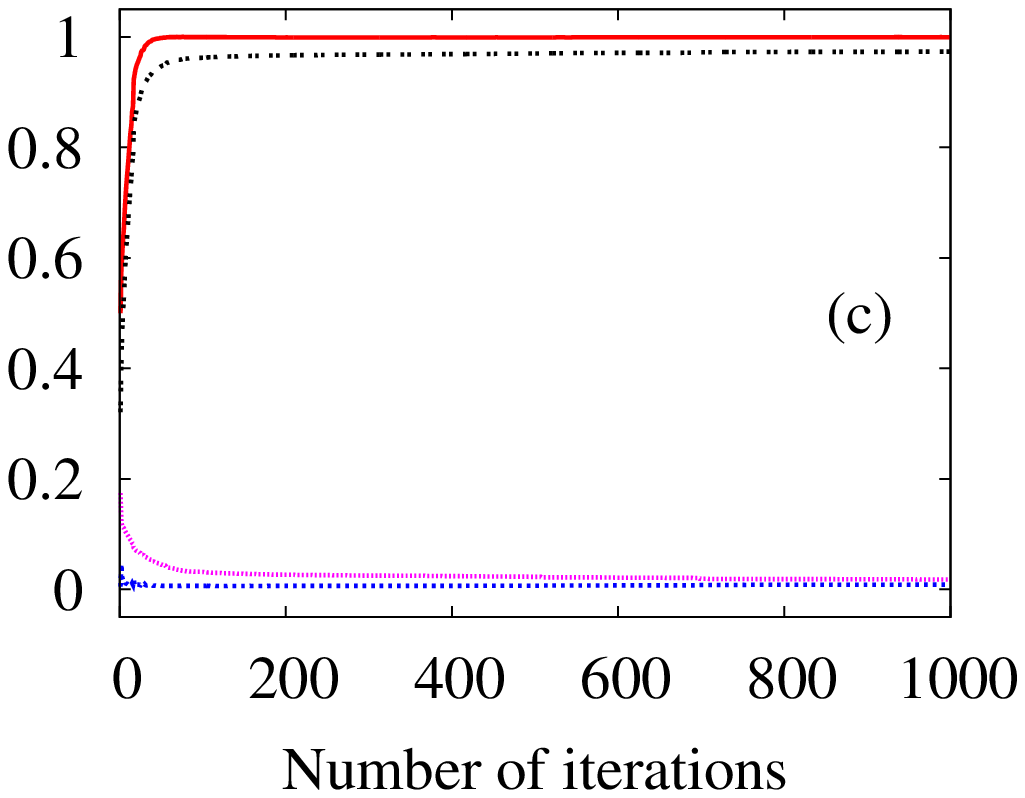} &
\includegraphics[width=0.285\textwidth,angle=0]{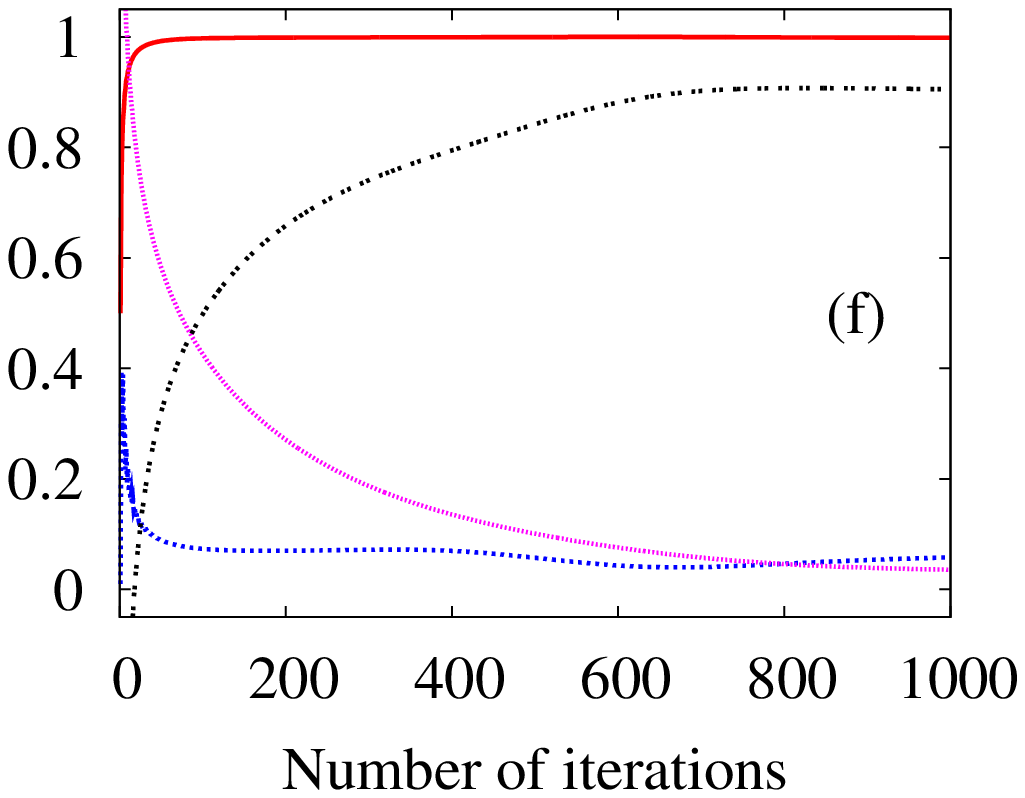} &
\includegraphics[width=0.285\textwidth,angle=0]{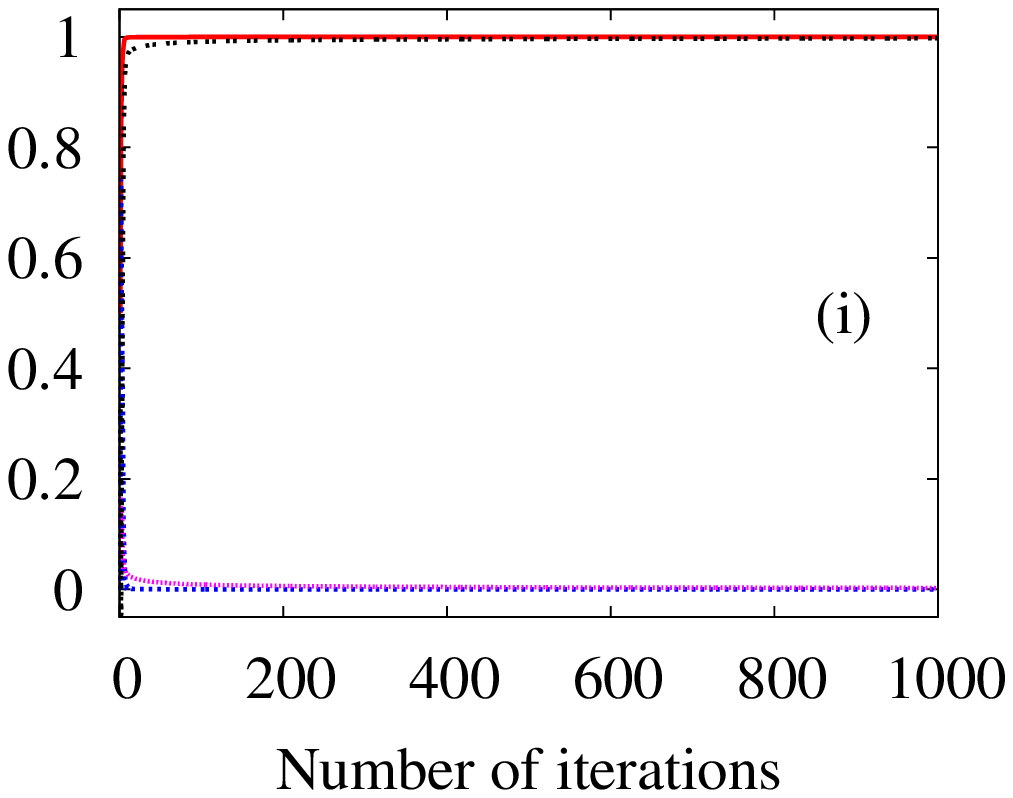} \\
\end{tabular}
\caption{Maximizing coherence in a three-level system using conjugate gradient method (a)-(c), Zhu-Rabitz method (d)-(f) and Krotov method (g)-(i).
(a), (d), (g) The population of state $\vert 1 \rangle$ - red line, state $\vert 2 \rangle$ - blue line, state $\vert 3 \rangle$ - black line and coherence $\varrho_{31}$ - pink line.
(b), (e), (h) The sequence of optimal pulses: Rabi frequency of the pump pulse - red line and the Stokes pulse - blue line.
(c), (f), (i) The convergence behavior of the optimized transition probability - red line, penalty on the field energy - blue line, penalty on the second state population - pink line and final optimized cost functional - black line versus the number of iterations.}\label{coherence}
\end{figure}

In this section we apply the OCT to create a maximum coherence $|\varrho_{31}|=|a_{3}^{*}a_{1}|$ between the levels $\vert 1 \rangle$ and $\vert 3 \rangle$ at final time $T$. We use the same parameters and the initial guess function for the fields envelopes as in the complete population transfer section. Figure~\ref{coherence} illustrates the results obtained using the conjugate gradient method (a)-(c), the Zhu-Rabitz method (d)-(f) and the Krotov method with $\Omega_{P,S}^r(t) \neq 0$ (g)-(i), respectively. This time we restrict our consideration to the case when a penalty on the intermediate state population is applied.

Figure~\ref{coherence}~(a), (d), (g) shows the time evolution of the population in the three-level system excited by the optimized pulse sequence. Dynamics of the population presented in the Fig.~\ref{coherence} is almost identical for the all three versions of the OCT: at the target time we obtain a maximum coherence $|\varrho_{31}(T)| \approx 1/2$, that is we create a 50/50 coherent superposition of states $|1\rangle$ and $|3\rangle$. Population of the intermediate state $|2\rangle$ is almost negligible during the excitation process.

The Rabi frequency of the pump and Stokes pulses obtained from the implementations of the OCT is shown in Fig~\ref{coherence}~(b), (e) and (h), respectively using the conjugate gradient method~\cite{Kis-2,balint-kurti,press}, the Zhu-Rabitz method~\cite{zhu} and the Krotov method with $\Omega_{P,S}^r(t) \neq 0$~\cite{palao}. All three methods give a similar solution for the optimal pulse sequence. The obtained pulse sequence and corresponding population dynamics allow us to conclude that the OCT finds the so-called half-STIRAP (also sometimes referred as fractional STIRAP) scheme as the optimal pulse sequence to create maximum coherence in three-level system. The same way as in the STIRAP scheme, the Stokes pulse is turned on first but both pulses are turned off simultaneously at the later time, see Fig~\ref{coherence}~(b), (e) and (h). The mechanism of the solution can be explained using the dressed state basis as follows. It is easy to find the eigenvalues and eigenvectors (dressed states) of the system Hamiltonian, Eq.~(\ref{eq2-interact}),  in the RWA. The important dressed state with energy $\lambda_{0}=0$ has the form
\begin{equation}\label{dark}
|c_{0}(t)\rangle = \Big\{\Omega_{S}^{}(t)/\sqrt{\Omega_{P}^{2}(t)+\Omega_{S}^{2}(t)}, 0, -\Omega_{P}^{}(t)/\sqrt{\Omega_{P}^{2}(t)+\Omega_{S}^{2}(t)} \Big\}  .
\end{equation}
That dressed state has zero projection on the pure state $|2\rangle$ while probability amplitudes to be in the state $|1\rangle$ and $|3\rangle$ are controllable by the ratio of the pump and Stokes Rabi frequencies. Analyzing  Eq.~(\ref{dark}) we can reproduce the population dynamics for the half-STIRAP scheme: at time $t=0$ the pump Rabi frequency $\Omega_{P}^{}(t=0)=0$ and the Stokes Rabi frequency $\Omega_{S}^{}(t)\neq 0$ therefore the dressed state $|c_{0}(t=0)\rangle$ correlates with the initial state $|1\rangle$; during the turn off stage of the pulse excitation ratio between pump and Stokes Rabi frequencies is equal to one, that results in creation of state $|c_{0}(t=T)\rangle = {1/\sqrt{2}, 0, -1/\sqrt{2}}$, the state of maximum coherence between states $|1\rangle$ and $|3\rangle$. Whole dynamics of the system takes place in one dressed states and that is possible only in the limit of the adiabatic regime when nonadibatic coupling to other two dressed states is negligible. According to our observation the OCT finds the adiabatic mechanism as an optimal solution of the control problem.

To compare the performance of the different implementations
the optimized transition probability ${\cal P}$, the final optimized cost functional $K$, the penalty on the field energy and the penalty on the second state population as a function of number of iterations are shown in Fig~\ref{coherence}~(c), (f) and (i). Best results for the maximal coherence obtained using the discussed methods are also summarized in Table~\ref{tab:table2}.

\begin{widetext}
\begin{table}[h]
\begin{center}
\caption{\label{tab:table2} Comparison of the results obtained using different implementation methods of the OCT to maximize coherence, $|\rho_{31}(T)|$, between levels $\vert 1 \rangle $ and $\vert 3 \rangle $ in the three-level $\Lambda$ system; $\varrho_{ii}(T)=|a_{i}(T)|^2$ is the population of state $\vert i \rangle$ , $i=1, 2, 3$. }
\vskip 5 pt
\begin{tabular}{|c|c|c|c|c|c|c|c|c|}
\hline
{\it Method}                 &  $\alpha_0$   &  $\beta$  &  $\varrho_{11}$  &  $\varrho_{22}$  &  $\varrho_{33}$  &  $|\varrho_{31}|$  &  ${\cal P}$  &  $K$  \\
\hline
{\bf Conjugate gradient}     &  0.00025      &  0.2  &  0.521  &  0.058  &  0.478  &  0.499   &  0.999   &  0.973   \\
\hline
{\bf Zhu-Rabitz}             &  0.0005       &  1.8  &  0.538  &  0.030  &  0.461  &  0.498   &  0.998   &  0.905    \\
\hline
{\bf Krotov ($\Omega_{P,S}^{r}\neq 0$)}                 &  0.1          &  0.2  &  0.503  &  0.011  &  0.496  &  0.499   &  0.999   &  0.997   \\
\hline
\end{tabular}
\end{center}
\end{table}
\end{widetext}

\section{Conclusion}
In this paper, we have analyzed a performance of several implementation methods of the OCT to design optimal pulse sequences for a complete population transfer and the creation of a maximum coherence in a three-level $\Lambda$ system. We have applied the conjugate gradient method~\cite{Kis-2,balint-kurti,press}, the Zhu-Rabitz method~\cite{zhu}, and the Krotov method~\cite{somloi,tannor,palao,palao1} to obtain the optimal solution.

Using the conjugate gradient method it was demonstrated earlier~\cite{malinovsky} that STIRAP type solution can
emerge automatically from the global OCT method. Now we have shown that the Zhu-Rabitz method~\cite{zhu} and the Krotov method~\cite{somloi,tannor,palao,palao1} provide an additional conformation that counterintuitive pulse sequence is the optimal solution according to the OCT. It was demonstrated that the penalty on the population of the intermediate state in a three-level system is a crucial factor to obtain the optimal STIRAP scheme.

We have also demonstrated that the half-STIRAP scheme is the optimal solution according to the optimal control theory for creation of a maximum coherence. We have shown that a creation of a maximally coherent superposition is one more optimization problem where the OCT finds the adiabatic mechanism as an optimal solution.

Note that the related results using OCT to create predetermined coherent superposition has been published recently~\cite{Kis-2}.
However, the target wave function used in~\cite{Kis-2} was different in comparison to present studies: population of the initially populated state at the later time was zero~\cite{Kis-2}. As the result, the generalized STIRAP pulse sequence was found as the optimal solution using the conjugate gradient method~\cite{Kis-2}. In our case, the optimization procedure reveals the half-STIRAP pulse sequence as the optimal solution
since we require that the half of the population should remain in the initially populated state. Comparing the results obtained using the above considered methods, it is clear that the methods can be applied to various problems on equal footing, since the value of optimized transition probability ${\cal P}$ is more than 99\% for all the methods. However, the value of optimized cost functional $K$ calculated using the conjugate gradient method is larger than corresponding values calculated using the Zhu-Rabitz iterative method and the Krotov method for $\Omega_{P,S}^{r}(t) = 0$ and $\beta \neq 0$. Note that the computation cost of the conjugate gradient method is generally larger than that of the other two methods.

One might observe that the pulses in optimal sequence obtained by  optimizing the population transfer (Figs.2-5) and the maximum coherence (Fig.6) have a relatively similar structure in all considered implementation schemes. However, the Zhu-Rabitz method~\cite{zhu} provides the shorter and more intense pulses in comparison to other methods. There is also a bit more pronounced asymmetry of the pump-pulse shape obtained by the Krotov method. We believe that these differences are due to the variations in the numerical implementation procedure and the values of the penalty parameters. The conjugate-gradient method~\cite{Kis-2,balint-kurti,press} incorporates one iteration to next iteration feedback from the control field while Krotov method~\cite{somloi,tannor,palao,palao1} incorporates one time step to next time step feedback of the control field, and Zhu-Rabitz method~\cite{zhu} incorporates feedback from the control field in an entangled fashion~\cite{Bonita}, mixing up both feedback techniques. The detailed description of the implementation procedure of these methods is given in the Appendices.

\section*{Acknowledgments}
The authors acknowledge a partial financial support from DARPA
HR0011-09-1-0008 and NSF PHY-0855391.

\appendix

\section{Conjugate gradient method}
For given target time $T$ and the number of time steps $N$ ($t_i = i \times \Delta t$, where $i=0,1,2,....,N$ and
$N\Delta t = T$), the conjugate-gradient method~\cite{Kis-2,balint-kurti,press} involves the following steps to obtain the optimal field:

{\bf Step 1:} Choose an initial electric field $\epsilon_{0}(t)$.

{\bf Step 2:} Set $k=1$ and $\epsilon^k(t)=\epsilon_{0}(t)$.

{\bf Step 3:} Propagate $|\psi^{k}(t=0)\rangle $ forward in time using $\epsilon^k(t)$ according to Eq.~(\ref{initial}) to obtain
$|\psi^{k}(T)\rangle$.

{\bf Step 4:} Evaluate the cost functional $K^k$ according to Eq.~(\ref{eq-K}). The last term in this equation is zero as $|\chi(t)\rangle$ is a solution of the time-dependent Schr\"odinger equation.

{\bf Step 5:} If $k\ge2$, compute $\Delta K^k=K^k-K^{k-1}$ and compare it with the convergence threshold, $\gamma$.
If $\Delta K^k \leq \gamma$, then stop the iteration and declare that the optimal pulse has been obtained.

{\bf Step 6:} Set $\langle \chi^k(T)|= \langle \phi(T)| \langle \phi(T) \vert \psi^k(T) \rangle$. Propagate $\langle \chi^k(T)|$ and $|\psi^k(T)\rangle$
backward in time using field $\epsilon^k(t)$ to obtain $\langle \chi^k(0)|$ and $|\psi^k(0)\rangle$.

{\bf Step 7:} The gradient $g^k(t)$ of the cost functional $K^k$ defined in Eq.~(\ref{eq-K}) with respect to the variation of
$\epsilon^k(t)$ at time $t$ is given by
\begin{equation}\label{A-1}
g^k(t) \equiv \frac{\partial K^k}{\partial \epsilon^k(t)} = -2 \Big[ \alpha(t) \epsilon^k(t) -
{\rm Im} \Big\langle \chi^k(t) \Big\vert \frac{\partial \hat{H}}{\partial \epsilon^k(t)} \Big\vert \psi^k(t) \Big\rangle \Big] \Delta t.
\end{equation}

The Polak-Ribiere-Polyak~\cite{A-polak} search direction is calculated using the Eq.~(\ref{A-1}) as
$$d^k(t) = g^k(t) + \zeta^k d^{k-1}(t) \,, $$
where
$$\zeta^k = \frac{g^k(t)^{T} \Big(g^k(t) - g^{k-1}(t)\Big)}{g^{k-1}(t)^{T}g^{k-1}(t)} \,,$$
$k$ = 2,3,..., $d^1(t) = g^1(t)$. The function $\zeta^k$ is the conjugate gradient update parameter and $g^k(t)^{T}$ is the transpose
of $g^k(t)$. A line search is then performed along this direction to determine the maximum value of the cost functional.

{\bf Step 8:} The electric field for the next $(k+1)$ iteration is taken as
$$\epsilon^{k+1}(t) = \epsilon^{k}(t) + \lambda \cdot d^k(t),$$
where $\lambda$ is determined by the line search, which makes $\epsilon^{k+1}(t)$ to generate the maximum value of $K$.

{\bf Step 9:} Go back to Step 3 and repeat 3-8 until the required convergence has been achieved.

\section{Zhu-Rabitz method}
Zhu-Rabitz method~\cite{zhu} involves the following steps to find the optimal value of the control field:

{\bf Step 1:} Choose an initial electric field $\epsilon_i(t)$. Set $k=1$ and $\epsilon^k(t)=\epsilon_{0}(t)$.

{\bf Step 2:} Propagate $|\psi^k(0)\rangle$ forward in time using $\epsilon^k(t)$ according to Eq.~(\ref{initial}) to obtain $|\psi^k(T)\rangle$.

{\bf Step 3:} Evaluate the cost functional $K^k$ according to Eq.~(\ref{eq-K}). If $k\ge2$, compute $\Delta K^k=K^k-K^{k-1}$ and compare it with the convergence threshold, $\gamma$. If $\Delta K^k \leq \gamma$, then stop the iteration and declare that the optimal pulse has been obtained.

{\bf Step 4:} Propagate Lagrange multiplier $\langle \chi^k(T)| = \langle \phi(T)| \langle \phi(T) \vert \psi^k(T) \rangle$ backward in time
using the new electric field $\epsilon^{k+1}(t)$ defined by
$$\epsilon^{k+1}(t) = \frac{1}{\alpha(t)} \cdot {\rm Im}
\Big\langle \chi^k(t) \Big\vert \frac{\partial \hat{H}}{\partial \epsilon^{k+1}(t)} \Big\vert \psi^k(t) \Big\rangle \,,
$$
according to Eq.~(\ref{lagrange}) to obtain $\langle \chi^k(0)|$.

{\bf Step 5:} Propagate $|\psi^{k+1}(0)\rangle$ forward in time using the new field $\epsilon^{k+1}(t)$ given by
$$\epsilon^{k+1}(t) = \frac{1}{\alpha(t)} \cdot {\rm Im}
\Big\langle \chi^k(t) \Big\vert \frac{\partial \hat{H}}{\partial \epsilon^{k+1}(t)} \Big\vert \psi^{k+1}(t) \Big\rangle \,,
$$
according to Eq.~(\ref{initial}) to obtain $|\psi^{k+1}(T)\rangle$.

{\bf Step 6:} Go back to Step 3 and repeat 3-5 until the required convergence has been achieved.

\section{Krotov method}
The Krotov method~\cite{somloi,tannor} employs a slightly different iteration procedure: \\

{\bf Step 1:} Choose an initial electric field $\epsilon_i(t)$.

{\bf Step 2:} Set $k=1$ and $\epsilon^k(t)=\epsilon_i(t)$.

{\bf Step 3:} Propagate $|\psi^k(t=0)\rangle$ forward in time with the field $\epsilon^k(t)$ according to Eq.~(\ref{initial}) to obtain $|\psi^k(T)\rangle$.

{\bf Step 4:} Evaluate the cost functional $K^k$ according to Eq.~(\ref{eq-K}). If $k\ge2$, compute $\Delta K^k=K^k-K^{k-1}$ and compare it with the convergence threshold, $\gamma$. If $\Delta K^k \leq \gamma$, then stop the iteration and declare that the optimal pulse has been obtained.

{\bf Step 5:} Set $\langle\chi^k(T)| = \langle\phi(T)| \langle \phi(T) \vert \psi^k(T) \rangle$. Propagate $\langle\chi^k(T)|$ backward in time
according to Eq.~(\ref{lagrange}) using field $\epsilon^k(t)$ to obtain $\langle\chi^k(0)|$.

{\bf Step 6a:} Set $|\psi^{k+1}(t=0)\rangle$ and propagate forward in time according to the Schr\"odinger equation~(\ref{initial}) with simultaneous evaluation of the electric field $\epsilon^{k+1}(t)$ at each time step. To calculate $\langle\chi(t)|$ use the old electric field, i.e. the electric field generated in the previous iteration step and calculate $|\psi^{k+1}(t)\rangle$ with the new electric field.

{\bf Step 6b:} Compute the new electric field $\epsilon^{k+1}(t)$ from $\langle\chi^k(t)|$ and $|\psi^{k+1}(t)\rangle$ according to
$$\epsilon^{k+1}(t) = \frac{1}{\alpha(t)} {\rm Im} \Big\langle \chi^k(t) \Big\vert \frac{\partial \hat{H}}{\partial \epsilon^{k+1}(t)}
\Big\vert \psi^{k+1}(t) \Big\rangle \,.$$

{\bf Step 7:} Go back to Step 4 and repeat 4-6 until the required convergence has been achieved.

\end{document}